\documentclass[aps,pra,twocolumn,superscriptaddress,floatfix]{revtex4-2}

\pdfoutput=1 
\usepackage[percent]{overpic}
\usepackage{graphicx,graphics}
\usepackage{dcolumn}
\usepackage{amsthm}
\usepackage{amsmath,amssymb,amsfonts}
\usepackage{latexsym,verbatim}
\usepackage{bm}
\usepackage{mathtools}
\usepackage{dsfont}
\usepackage{bbold}
\usepackage{color}
\usepackage{ulem}
\usepackage[breaklinks=false,colorlinks,citecolor=blue,linkcolor=blue,urlcolor=blue]{hyperref}

\usepackage{braket}
\usepackage{soul}
\begin{document}
\title{Optimal Control Methods for Quantum Batteries}

\author{Francesco Mazzoncini}
\email{mazzoncini@telecom-paris.fr}
\affiliation{ T\'el\'ecom Paris-LTCI, Institut Polytechnique de Paris, 19 Place Marguerite Perey, 91120 Palaiseau, France}

\affiliation{NEST, Scuola Normale Superiore, I-56126 Pisa, Italy}

\author{Vasco Cavina}
\affiliation{NEST, Scuola Normale Superiore, I-56126 Pisa, Italy}
\affiliation{Department of Physics and Materials Science, University of Luxembourg, L-1511 Luxembourg, Luxembourg}

\author{Gian Marcello Andolina}
\affiliation{NEST, Scuola Normale Superiore, I-56126 Pisa, Italy}
\affiliation{ICFO-Institut de Ci\`{e}ncies Fot\`{o}niques, The Barcelona Institute of Science and Technology, Av. Carl Friedrich Gauss 3, 08860 Castelldefels (Barcelona),~Spain}

\author{Paolo Andrea Erdman}
\affiliation{Freie Universit{\" a}t Berlin, Department of Mathematics and Computer Science, Arnimallee 6, 14195 Berlin, Germany}

\author{Vittorio Giovannetti}
\affiliation{NEST, Scuola Normale Superiore, I-56126 Pisa, Italy}

\date{\today}

\begin{abstract}
We investigate the optimal charging processes for several models of quantum batteries, finding how to maximize the energy stored in a given battery with a finite-time modulation of a set of external fields. We approach the problem using advanced tools of optimal control theory, highlighting the universality of some features of the optimal solutions, for instance the emergence of the well-known Bang-Bang behavior of the time-dependent external fields. The technique presented here is general, and we apply it to specific cases in which the energy is both pumped into the battery by external forces (direct charging) or transferred into it from an external charger (mediated charging). In this article we focus on particular systems that consist of coupled qubits and harmonic oscillators, for which the optimal charging problem can be explicitly solved using a combined analytical-numerical approach based on our optimal control techniques. However, our approach can be applied to more complex setups, thus fostering the study of many-body effects in the charging process.
\end{abstract}
\maketitle

\section{Introduction}

In recent years, with the rapid development of new quantum technologies \cite{QuantumTechnologiesProgramme1, QuantumTechnologiesProgramme2}, there has been a worldwide interest in exploiting quantum phenomena that arise at a microscopic level. Here, we will focus on studying the so-called ''quantum batteries"\cite{Qbat_Alicki,Qbattery1,Qbattery2,Qbattery3,Charging1,Charging2,Charging3,Charging4}, i.e. quantum mechanical systems employed for energy storage, where quantum effects can be used to obtain more efficient and faster charging processes than classical systems.

This blossoming research field has to address many different questions, such as the stabilization of stored energy \cite{Gherardini2020,Rosa20}, the practical implementation of quantum batteries \cite{ferraro,Quach2022}, and the study of the optimal charging processes \cite{Binder2015a, Campaioli2017,Rodriguez2022}, offering a vast research panorama on both theoretical \cite{Ferraro2018,Pirmoradian2019,ferraro,Barra2019,Monsel2019,Rossini,Gherardini2020,Rosa20,Rossini2020,Mitchison2021,Hovhannisyan20,Gyhm2022,Barra2022} and experimental ends \cite{Quach2022,hu2021}.
Within this framework, we will derive optimal charging strategies for quantum batteries using techniques from Quantum  Control Theory
\cite{Qcontrolreview1, Qcontrolreview2, Qcontrolreview3},
a powerful mathematical tool that has many applications in different fields of physics such as quantum optics \cite{Qcontrolopt1} and physical chemistry \cite{Qchim1, Qchim2, Qchim3}.
Quantum control theory has contributed to understanding interesting aspects of quantum mechanics such as the quantum speed limit \cite{QSL1, QSL2, QSL3, QSL4} and to generate efficient quantum gates in open quantum systems \cite{ControlOpen1, ControlOpen2}.
In this work, we  study how a qubit  or a quantum harmonic oscillator  can be optimally charged with a modulation of an external Hamiltonian. In order to find the best charging protocol we will use the Pontryagin's Minimum Principle (PMP) \cite{Pontryagin, KirkOptimal}, a very useful theorem of  Classical Optimal Control Theory, which is frequently used also in Quantum Control Theory \cite{Optimalcontrol1, Optimalcontrol2}. We show that, in most cases that we consider,  quantum batteries can be optimally charged through different variants of a so called  \textit{Bang-Bang} modulation  of the intensity of an external Hamiltonian.

Our paper is organized as follows. In section \ref{section:general_charging} we introduce two general  charging protocols to inject energy in a quantum battery. In section \ref{section:PMP} we present a brief introduction  to Pontryagin's Minimum Principle, highlighting the main tools that we shall use throughout  the paper.
In section \ref{section:onesystem} we focus on the first charging  protocol, consisting  of a closed system charged by the modulation of an external Hamiltonian. Section \ref{section:charger-mediated}
is devoted to analyzing a second charging process, where we make use of the coupling between a quantum battery and an auxiliary quantum system. Finally, a brief summary of our main conclusions is reported in section \ref{section:conclusion}, while useful technical details can be found in the appendix.

\section{Charging of a Quantum Battery}
\label{section:general_charging}
\begin{figure}[t!]
\centering
\centering
\begin{overpic}[width=0.95\columnwidth]{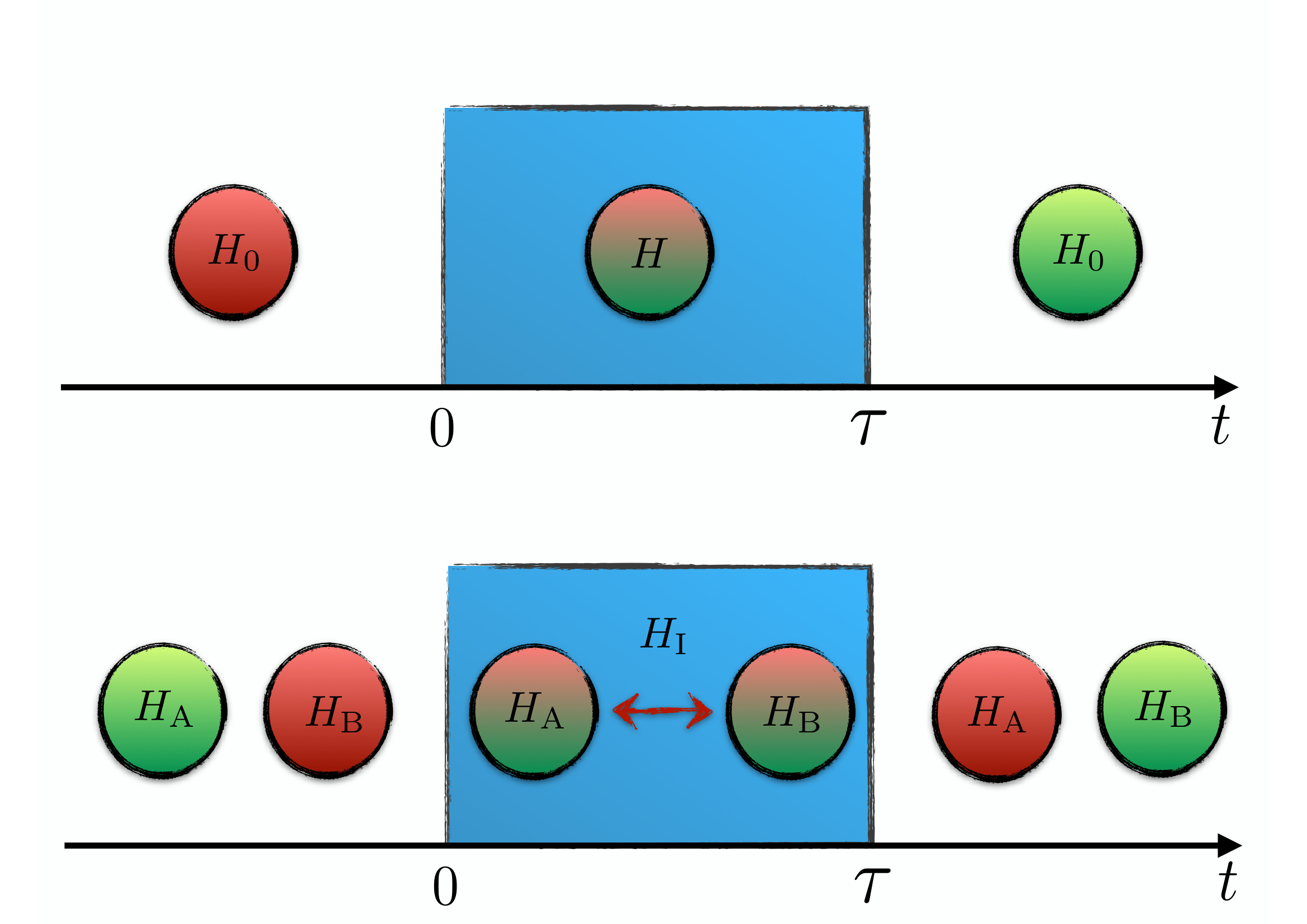}
\put(0,60){(a)}\put(0,30)    {(b)}
\end{overpic}
      
\caption{a) Direct Charging Process: charging model for a closed system through the modulation of an external control  for a finite amount of time $\tau$. b) Mediated Charging Process: it consists in letting two systems A and B interact through an Hamiltonian $H_1$.}
\label{fig:charging_model}
\end{figure}

We start defining two general  protocols for the charging process of a quantum battery, see Fig.~\ref{fig:charging_model} for a pictorial representation.

\paragraph{Direct Charging Process (DCP):} 
The first charging model consists of a single closed quantum system initialized in a state $\rho(0)$ that evolves in time under the action of a time-dependent Hamiltonian of the form 
 \begin{equation}
\label{eq:charging_model}
H(t)= H_0 +{\bm{\lambda}}(t) \cdot {\mathbf H} := H_0 +
\sum_{i=1}^m \lambda_i (t) H_i\;.
\end{equation}
In this expression $H_0$ is the {\it intrinsic} Hamiltonian contribution which defines the energy content of the system before and after the charging process and ${\mathbf H}:= (H_1,\cdots, H_m)$ is a collection of 
 {\it charging} Hamiltonian terms  which are modulated by control functions 
${\bm{\lambda}}(t) := (\lambda_1(t), \cdots,\lambda_m(t))$
that we assume to be active (i.e. different from zero) only over a limited time interval $[0,\tau]$. They can take values that are determined by some assigned constraint, i.e. ${\bm{\lambda}}(t) \in {\mathbb D}[0,\tau]$,
where
 $\tau>0$ is the total duration of the charging process and ${\mathbb D}[0,\tau]$ is a proper subset of the real functions ${\mathbb F}[0,\tau]$
 mapping $[0,\tau]$ into ${\mathbb R}^{m}$.
Our goal is hence to find an optimal ${\bm{\lambda}}^{\star}(t) \in {\mathbb D}[0,\tau]$ that, given an assigned $\tau$, maximizes the mean energy of the system at the end of the process.
Introducing
\begin{equation} 
U_{\tau} :={\cal T}\exp[-i \int_0^{\tau} dt~ H(t)]\;,
\label{dfds}
\end{equation}
 the time-ordered unitary evolution operator associated with the
time-dependent Hamiltonian~(\ref{eq:charging_model}), and  
\begin{eqnarray} 
\rho(\tau)=U_{\tau} \rho(0) U^{\dag}_{\tau}\;, \label{unitev}
\end{eqnarray} 
the evolved state of the system at time $\tau$, 
 we aim to determine the quantity
\begin{eqnarray} \label{max}
 E_{\max}(\tau)  := E(\tau)\Big|_{{\bm{\lambda}}^{\star}(t)} = \max_{{\bm{\lambda}}(t)\in {\mathbb D}[0,\tau]} E(\tau) \;,
\end{eqnarray}  
where  using $\langle \,\cdot\, \rangle$  as a short-hand notation to indicate the trace operator, we set
\begin{eqnarray}
E(\tau)  := \langle\rho(\tau) H_0 \rangle \;, \label{outmean} 
\end{eqnarray}
(notice that hereafter we have set $\hbar=1$).
It is worth pointing out that since the DCP models
 considered here rely on closed dynamical evolutions  (no interactions with  external degrees of freedom being allowed),
the DCP optimization we are targeting corresponds also to maximizing the amount of {\it extractable work} we can store in the system as measured
by the ergotropy, the total ergotropy, or the thermal free-energy ~\cite{Niedenzu2019}. To see this explicitly we recall that given a quantum system with Hamiltonian $H(t)$ and state $\rho(t)$, all these quantities can be computed as 
\begin{eqnarray}\label{dfdf} 
\mathcal{W}[\rho(t), H(t)]&:=& \langle \rho(t)  H(t) \rangle - {\cal F}(s_{\rho(t)},s_{H(t)})\;,
\end{eqnarray} 
where  ${\cal F}(s_{\rho(t)},s_{H(t)})$ is a functional that only depends upon the collections $s_{\rho(t)}=\{ \eta_1(t), \eta_2(t), \cdots\}$ and $s_{H(t)}=\{ \epsilon_1(t), \epsilon_2(t), \cdots\}$
of the eigenvalues of $\rho$ and $H$ respectively (see App.~\ref{app:defergo} for details). 
Since the unitary evolution \eqref{unitev} preserves $s_{\rho(t)}$, and $s_{H(
\tau)} = s_{H_0}$ in the DCP,  ${\cal F}(s_{\rho(\tau)},s_{H(\tau)}) = {\cal F}(s_{\rho(0)},s_{H_0})$ so that this quantity plays no role in the optimization procedure.

\paragraph{Mediated Charging Process (MCP):}  Although the DCP is of undoubted theoretical interest,  a closed evolution of a unique system is not genuinely realistic from the physical implementation's point of view. Such unitary evolution regime occurs only when the dynamics of the energy source are very slow compared to the Quantum Battery dynamics (i.e.\ in the Born-Oppenheimer limit).
Therefore, we also consider a second charging model, called \textit{charger-mediated process}~
\cite{Qbattery1,Qbattery2}, that involves instead two separate elements: an auxiliary quantum system A, called charger, and a quantum battery B. 
In the MCP we aim at maximizing the energy stored in B by suitably modulating its interaction with A in finite time $\tau$. 
For this sake we replace the DCP hamiltonian \eqref{eq:charging_model} with
\begin{equation} \label{eq:nuovach} 
 H(t):= H_A + H_B + {\bm{\lambda}}(t) \cdot {\mathbf H}\;, 
\end{equation} 
where $H_A$, $H_B$ are local operators of $A$ and $B$ respectively and $\boldsymbol{H}$ is now free to act on both the battery and the auxiliary system.
The quantity to optimize is now given by 
\begin{eqnarray} \label{MCPmerit}
 E_B(\tau):=\langle \rho_B(\tau)H_B\rangle \;, \end{eqnarray} 
where $\rho_B(\tau)$ is the reduced density matrix of the battery at time $\tau$.
Since $\rho_B(\tau)$  does not follow a unitary trajectory in the MCP scenario,  $s_{\rho_B(\tau)}$ is typically different from  $s_{\rho_B(0)}$, this implies that  $\mathcal{W}[\rho_B(\tau), H_B]$ 
is considerably more challenging to optimize.  We shall see however that  by choosing wisely the global initial state $\rho_{AB}(0)$, we can reduce our analysis to simpler DCPs, as shown in Sec.~\ref{section:charger-mediated}.

\section{Pontryiagin's Minimum Principle}
\label{section:PMP}

The Pontryiagin's Minimum Principle (PMP) \cite{KirkOptimal} is the main tool we will use in the optimization of DCPs and MCPs and will allow us to formally identify necessary conditions for the optimality of ${\bm{\lambda}}^{\star}(t)$. 
Here we introduce the approach to optimal control problems provided by PMP using a general formalism, that will be adapted to both DCP and MCP problems afterwards.
Consider a set of \textit{state variables} at a given time $t$, represented by the elements of a vector $\bm{v}(t):= (v_1(t), \cdots, v_n(t))$  which evolves 
via a dynamical equation  represented by  $n$ first-order differential equations of the form 
\begin{equation}
\dot{\bm{v}}(t)= \bm{f}(\bm{v}(t),{\bm{\lambda}}(t),t)\;,
\end{equation}
with $\bm{f}$ a vectorial function. The quantity to optimize, also called
{\it  perfomance criterion} is evaluated  in terms of a cost function written as 
 
\begin{equation}
J=\int^{\tau}_{0} g(\bm{v}(t), {\bm{\lambda}}(t),t)dt,
\label{functional}
\end{equation}
with $g$  a scalar function. Defining the \textit{pseudo-Hamiltonian} $\mathcal{H}$ as
\begin{equation}
\mathcal{H}:= g(\bm{v}(t), {\bm{\lambda}}(t), t) + {\bm{p}}(t)\cdot {\bm{f}}({\bm{v}}(t),{\bm{\lambda}}(t), t),
\label{pseudohamgen}
\end{equation}
with ${\bm{p}}(t)$ the $n$-dimensional row vector of Lagrange multipliers, called {\it costates},
the PMP states that  necessary conditions for an optimal control ${\bm{\lambda}}^{\star}(t)\in \mathbb{D}[0,\tau]$  to minimize $J$ are  that for all $t \in [0,\tau]$:
\begin{equation}
\begin{cases}
 \dot{{\bm{v}}}(t)= \frac{\partial \mathcal{H}}{\partial {\bm{p}}}({\bm{v}}(t),{\bm{\lambda}}^{\star}(t), {\bm{p}}(t),t)\;,\\\\
 \dot{{\bm{p}}}(t)= - \frac{\partial \mathcal{H}}{\partial {\bm{v}}}({\bm{v}}(t),{\bm{\lambda}}^{\star}(t), {\bm{p}}(t),t)\;,\\\\
 \mathcal{H}({\bm{v}}(t), {\bm{\lambda}}^{\star}(t), {\bm{p}}(t),t) \leq \mathcal{H}({\bm{v}}(t), {\bm{\lambda}}(t), {\bm{p}}(t),t)\;,   \\ \qquad \qquad \qquad \qquad \qquad \qquad \qquad  \forall {\bm{\lambda}}(t)\in \mathbb{D}[0,\tau]\;.
\end{cases}
\label{cond4}
\end{equation}
Moreover, the PMP gives additional constraints based on the boundary conditions of our problem, i.e. whether the final state and the final time are fixed or free. In particular: \begin{itemize}
\item if the  final time $\tau$ is fixed and  no constraint  is posed on the final state ${\bm{v}}(\tau)$, then
\begin{equation}
{\bm{p}}(\tau)=(0, 0,\cdots,0)\;;
\label{statefree}
\end{equation}
\item  if the final time $\tau$ is free while the final state ${\bm{v}}(\tau)$ is fixed, then 
\begin{equation}
\mathcal{H}({\bm{v}}(\tau), {\bm{\lambda}}^{\star}(\tau),{\bm{p}}(\tau), \tau) =0\;.
\label{taufree}
\end{equation}
\end{itemize}
We finally highlight that the  PMP is not the only possible optimization method to  analyze charging processes for quantum batteries.  For instance, Ref. \cite{Rodriguez2022} deploys an iterative approach to minimize the distance between the target state and the final state, considering a variant of our charger-mediated process where a field is modulated only acting on the charger, considered in this case as an open dissipative system. However, since PMP gives necessary conditions for optimality, any other optimization method must eventually satisfy those conditions.

\section{DCP Optimal solutions}\label{section:onesystem}
In this section we will derive the optimal solutions for DCPs considering two different settings:
first in Sec.~\ref{PMPenergy}  we fix the total duration of the charging event $\tau$ and try to identify the optimal pulse $\bm{\lambda}^\star(t)$ which, starting from a given initial configuration $\rho(0)$,  produces the maximum value of the output energy $E_{\max}(\tau)$; then in Sec.~\ref{PMPtime} we analyze the opposite problem: that is we fix a target output state that ensures a certain 
value of the final energy, and try find the optimal control $\bm{\lambda}^\star(t)$  that enable us to reach it 
in the minimum time~$\tau$.

\subsection{Maximum output energy at fixed time \texorpdfstring{$\tau$}{Lg}} \label{PMPenergy} 
 To begin with, we observe that if i) the charging Hamiltonian terms $H_i$'s are generators of the group 
${\cal U}$ of the unitary operators on the system,  and  ii) no restrictions are imposed on the choice of the control vector ${\bm{\lambda}}(t)$, allowing ${\mathbb D}[0,\tau]$ to include all possible elements
${\mathbb F}[0,\tau]$, then the dynamical evolutions (\ref{dfds}) can span  the entire set ${\cal U}$ of unitary transformation on the system. Accordingly under conditions i) and ii) we can write  
\begin{eqnarray}\!\!\!\! E_{\max}(\tau)&=&
\max_{{\bm{\lambda}}(t)\in {\mathbb F}[0,\tau]}\langle U_\tau \rho(0) U_\tau^\dag H_0\rangle  \nonumber \\
&=& \max_{U\in {\cal U}} \langle U\rho(0) U^\dag H_0\rangle=: \overline{E}_{\max}  \;, \label{anti0} 
\end{eqnarray} 
where $\overline{E}_{\max}$ is a $\tau$ independent constant
that represents the maximum amount of energy we can force into the system via arbitrary unitary manipulations.
The constant $\overline{E}_{\max}$ can be explicitly evaluated as 
 \begin{equation}
\overline{E}_{\max} \label{anti}
 =\sum_{i=1} \eta^{\uparrow}_i(0) \; \epsilon_i^{\uparrow}(0)\;,
 \end{equation} 
 with 
 $s^{\uparrow}_{\rho(0)}\!\!=\!\{ \eta^{\uparrow}_1\!(0), \eta^{\uparrow}_2 \!(0), \dots\}$ and $s^{\uparrow}_{H_0}\!\!\!=\!\{ \epsilon^{\uparrow}_1\!(0), \epsilon^{\uparrow}_2\!(0),\dots  \}$ being 
the spectra
of $\rho(0)$ and $H_0$, rearranged in increasing order.
Note that it is possible to establish a direct connection between  $\overline{E}_{\max}$ and the anti-ergotropy \cite{Salvia2021} of the system (see App. \ref{def:ergo} for details).
Apart from this special case, the explicit evaluation of $E_{\max}(\tau)$ is typically rather demanding and does not admit a closed analytical solution. One possible approach to tackle it is to make use of optimal control techniques. In particular, in what follows we shall rely on the  
 PMP we have reviewed in Sec.~\ref{section:PMP}. For this purpose we  rewrite the final energy~(\ref{outmean}) as
\begin{equation}
E(\tau)=\int^{\tau}_0 \bigr \langle H_0 \dot{\rho}(t)\bigr \rangle dt  + E(0)\;,
\label{en2}
\end{equation}
where $\dot{\rho}(t) = \mathcal{N}[\rho(t)]=-i[H(t), \rho(t)]$. 
Accordingly, we can study the optimization of the charging process as a minimization problem of the cost function 
\begin{equation} \label{THECOST} 
J:= -\int^{\tau}_0 \bigr \langle H_0 \mathcal{N}[\rho(t)]\bigr \rangle dt\;.
\end{equation}
The optimization task can then be translated into a  PMP problem by introducing  the following arrangements
\begin{align}
\label{eq:PMPtoQuantum} \notag
& {\bm{v}}(t) \rightarrow \rho (t) \;, \quad \quad {\bm{\lambda}}(t) \rightarrow {\bm{\lambda}} (t) \;,  \quad \quad {\bm{p}}(t)  \rightarrow \pi (t) \;,
 \\ \notag
& {\bm{f}}({\bm{v}}(t) ,{\bm{\lambda}}(t),t) \rightarrow \mathcal{N} [\rho(t)]\;, \\ \notag
& g({\bm{v}}(t),  {\bm{\lambda}}(t),t) \rightarrow -\langle H_0 \mathcal{N}[\rho(t)]\bigr \rangle\;,  \\
& {\bm{p}}(t)\cdot {\bm{f}}( \bm{x}(t),{\bm{\lambda}}(t), t) \rightarrow \langle \pi(t) \mathcal{N}[\rho(t)]\rangle\;, 
\end{align}
with $\pi(t)$ being a self-adjoint operator of the same dimension of $\rho(t)$,
and defining the pseudo-Hamiltonian
\begin{eqnarray} \label{defiH} 
\mathcal{H}(\rho(t),{\bm{\lambda}}(t),\pi(t),t) &:=& \langle ( \pi(t)-H_0){\cal N}[\rho(t)]\rangle \\
&=&{\bm{\lambda}}(t) \cdot {\bm{G}}(t)-i\langle \pi'(t) [H_0,\rho(t)]\rangle\;, \nonumber 
\end{eqnarray}
 with $\pi'(t) := \pi(t)-H_0$ and ${\bm{G}}(t):=(G_1(t), \cdots, G_m(t))$ being a column-vector of elements \begin{eqnarray}
 G_j(t):=-i\langle\pi'(t)[H_j,\rho(t)]\rangle\;. \label{defGj} 
 \end{eqnarray} 
This allows us to express the necessary conditions~\eqref{cond4} for the optimal control vector ${\bm{\lambda}}^{\star}(t)$ as
\begin{equation}
\begin{cases}
\dot{\rho}(t)= -i\big[ H^{\star}(t), \rho(t)\big]\;, \\ \\
 \dot{\pi}'(t)=-i[H^{\star}(t),\pi'(t)]\;, \\ \\
  {\bm{\lambda}}^{\star}(t) \cdot {\bm{G}}(t) \leq  {\bm{\lambda}}(t) \cdot {\bm{G}}(t)\;, 
\qquad \quad  \forall{\bm{\lambda}}(t)\in \mathbb{D}[0,\tau]\;, 
\end{cases}
\label{condizionifin}
\end{equation}
where  $H^{\star}(t)$  represents the Hamiltonian~(\ref{eq:charging_model}) evaluated on the optimal control pulse, i.e. 
  \begin{equation} H^{\star}(t):=H_0+{\bm{\lambda}}^{\star}(t)\cdot {\mathbf{H}}\;. \label{HAM*} \end{equation} 
 In the third line of Eq.~(\ref{condizionifin}) we exploited Eq.~(\ref{defiH}) and the fact that  
 the term $-i\langle\pi'(t) [H_0,\rho(t)]\rangle$ does not depend explicitly on ${\bm{\lambda}}(t)$.
In the case of  a charging process with fixed time $\tau$ and unknown optimal final state $\rho(\tau)$, the list~(\ref{condizionifin}) has to be completed with the extra condition~\eqref{statefree} which in the present case becomes  \begin{eqnarray}\pi(\tau)=0 \quad \Longleftrightarrow \quad \pi'(\tau)=-H_0\;. \label{extra} \end{eqnarray} 
The first  two equations in~(\ref{condizionifin}) simply tell us that $\rho(t)$ and $\pi'(t)$ represent the state and the costate operator of the system evolved under the action of the Hamiltonian~(\ref{HAM*}).
What ultimately decides whether a given $\bm{\lambda}^\star(t)$ has a chance of being an
optimal solution is the inequality in Eq.~(\ref{condizionifin}) which, unfortunately, due to the implicit dependence upon $\bm{\lambda}^\star(t)$ of ${\bm{G}}(t)$, is typically  not analytically treatable. Nonetheless, in the special  special case where we have a unique control function (i.e. $m=1$) and the allowed domain $\mathbb{D}[0,\tau]$ is chosen to  simply force the intensity of $\lambda_1(t)$ to belong to a given interval ${\cal I}_1 = [\lambda_1^{\min}, \lambda_2^{\max}]$,
 the
inequality in Eq.~(\ref{condizionifin}) translates into  a series of  (simplified) conditions 
which provide us with a  nice guidance on how to construct the optimal 
control pulse, i.e.
\begin{itemize}
\item[a)] $\lambda_1^{\star}(t)$ can take the {\it minimum}  allowed value $\lambda_1^{\min}$ iff  the associated $G_1(t)$ function is strictly positive, i.e. 
$$\lambda_1^{\star}(t)=\lambda_1^{\min} \, \, \, \Longleftrightarrow   \, \, \, G_1(t)>0 \;;$$
\item[b)] $\lambda_1^{\star}(t)$ can take the {\it maximum} allowed value $\lambda_1^{\max}$  iff  the associated $G_1(t)$ function is strictly negative, i.e. 
$$\lambda_1^{\star}(t)=\lambda_1^{\max} \, \, \, \Longleftrightarrow \, \, \, G_1(t)<0 \;;$$
\item[c)] $\lambda_1^{\star}(t)$ can take arbitrary values in the allowed domain ${\cal I}_1:=[\lambda_1^{\min}, \lambda_1^{\max}]$
iff the associated $G_1(t)$ is equal to zero.
\end{itemize} 
\begin{figure}[t!]
\centering
\includegraphics[width= 0.45 \textwidth]{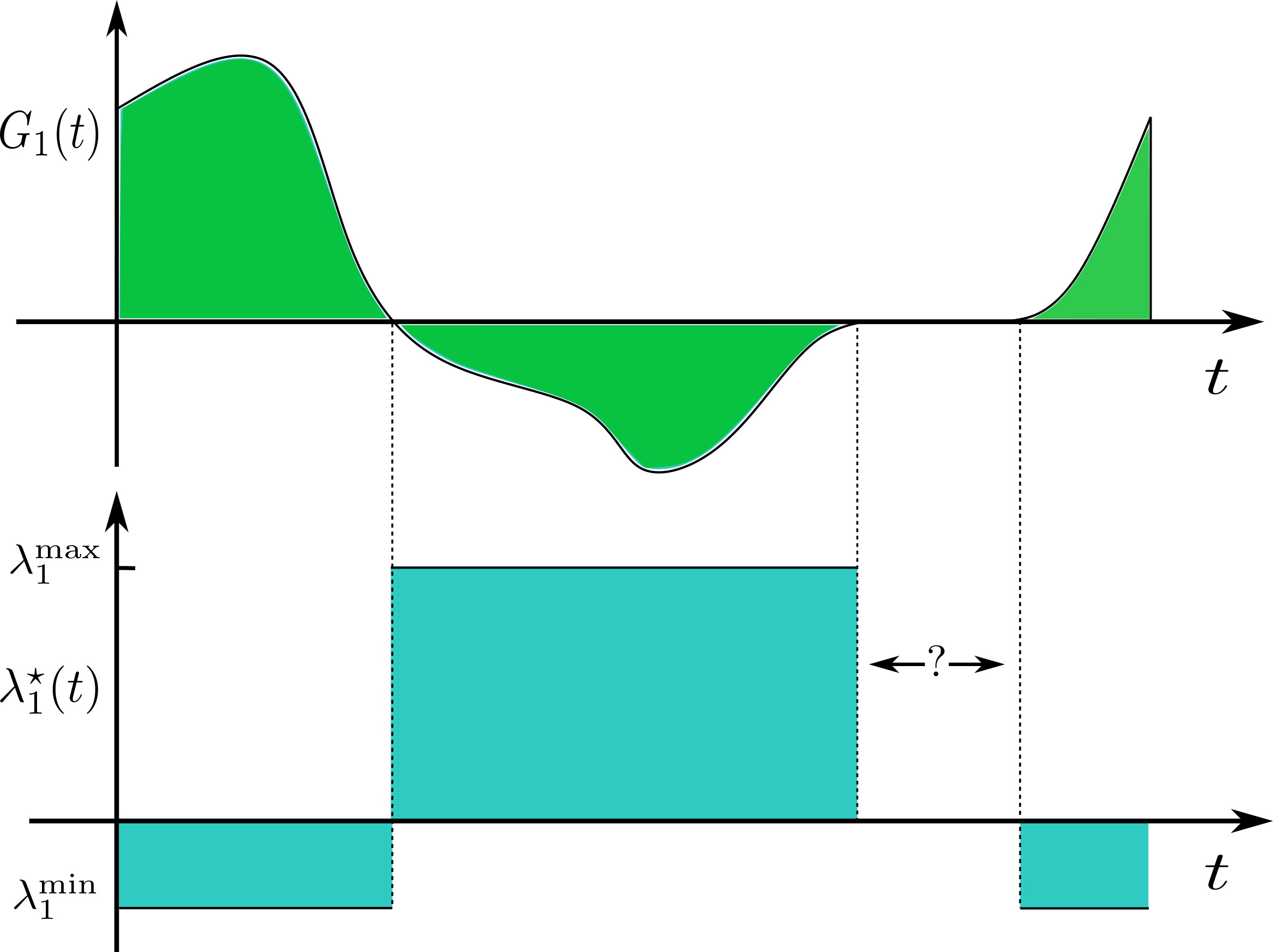}
\caption{Example of the relationship between a time-optimal control  $\lambda_1^{\star}(t)$ and the $G_1(t)$ function for the case in which the system is characterized by a single control function ($m=1$).  The region with a question mark is a singular interval, where the value of the optimal control is not determined by the conditions in Eq.~(\ref{condizionifin}).}
\label{exampBang}
\end{figure}
From the above analysis 
it emerges that natural candidates for  $\lambda_1^{\star}(t)$ are {\it Bang-Bang}-like step functions similar to the one shown in Fig.~\ref{exampBang} which alternate their values among 
the allowed extreme  $\lambda_1^{\min}$ and $\lambda_1^{\max}$ with switching points corresponding to the zeros of the associated $G_1(t)$ function.
The only allowed exceptions to this rule is when  
 $G_1(t)$  is zero over an extended interval ({\it singular interval} scenario):  in this case the necessary conditions in Eq.~(\ref{condizionifin}) give no information about how to select $\lambda_1^{\star}(t)$ 
without specifying the nature of the system. 

\subsection{Minimum charging time at fixed final state}\label{PMPtime} 

 Another problem that we can tackle using  the PMP method is to determine the minimum value of 
the charging time $\tau$ that allows us to move our initial state $\rho(0)$ into a final target configuration
$\rho_{\diamond}$ --- for instance 
a state in which the eigenvalues are sorted in increasing order
which according to Eq.~(\ref{anti})
grants us the maximum value of the stored final energy $\overline{E}_{\max}$ allowed by the most general DCP 
process.  
The new  cost function of the problem can be written as
 \begin{equation}
 J:=- \int^{\tau}_0 1 dt\;,
 \end{equation}
which is a simple way to express in an integral form the charging time.
 With the same notations adopted in the previous section, we can hence define the pseudo-Hamiltonian
 of the new problem as
 \begin{eqnarray}
 &&\mathcal{H}(\rho(t),{\bm{\lambda}}(t),\pi(t),t):= \langle\pi(t) (\mathcal{N}[\rho(t)])\rangle -1 \nonumber \\
 &&\qquad \qquad \qquad ={\bm{\lambda}}(t) \cdot {\bm{G}}(t)  -i\langle(\pi(t) [H_0,\rho(t)]\rangle - 1\;, \nonumber 
 \label{pseudoham_time}
 \end{eqnarray}
 where now ${\bm{G}}(t)$ is the vector of components
 \begin{eqnarray}\label{defGnuovo} 
 G_j(t)=-i\langle\pi(t) [H_j,\rho(t)]\rangle\;.
 \end{eqnarray}
 Doing almost the same calculations that led us to Eq.~(\ref{condizionifin}), we can hence cast the PMP
 constraint~\eqref{cond4}  for the optimal pulse $\bm{\lambda}^\star(t)$ that leads
 to the target state $\rho_{\diamond}$ in the minimal time $\tau$, in the following form 
 \begin{equation}
\begin{cases}
\dot{\rho}(t)= -i\big[ H^{\star}(t), \rho(t)\big]\;, \\ \\
 \dot{\pi}(t)=-i[H^{\star}(t),\pi(t)]\;, \\ \\
  {\bm{\lambda}}^{\star}(t) \cdot {\bm{G}}(t) \leq  {\bm{\lambda}}(t) \cdot {\bm{G}}(t)\;, 
\qquad \quad  \forall{\bm{\lambda}}(t)\in \mathbb{D}[0,\tau]\;, 
\end{cases}
\label{condizionifinNEW}
\end{equation}
 where  $H^{\star}(t)$  is again defined as in (\ref{HAM*}),
with the new extra condition
 imposed by~(\ref{taufree}) 
 \begin{eqnarray} 
 \langle\pi(\tau)[H^{\star}(\tau),\rho_{\diamond}]\rangle=-i \;, \label{newextra} 
 \end{eqnarray} 
 replacing Eq.~(\ref{extra}). Notice that also in this case $\rho(t)$ and the costate $\pi(t)$ undergo the same temporal dynamics; 
 however, in the present problem the final value of the costate is only partially determined by the new constraint Eq.~(\ref{newextra}). We also point out that as  for (\ref{condizionifin}) simplifications arise when there is only one control parameter $m=1$ with constrained intensity $\lambda_1(t) \in {\cal I}_1$, which allows one to translate the third equation of~(\ref{condizionifinNEW}) into the same
 a), b), c) rules detailed in the previous section.

 \section{Examples of DCP models} \label{EXAMPLES} 
 Here we analyze in details some examples of DCP models: a qubit with one ($m=1$) or two ($m=2$) charging fields, and an harmonic oscillator under the action of a linear, time-dependent perturbation.

\subsection{Qubit Optimal DCP with a single charging field}
\label{qubit_optimal}
In this section we focus on a first example of DCP where the system of interest is represented by a single qubit which is controlled via a single control field (i.e. $m=1$). For the Hamiltonian~(\ref{eq:charging_model}) we select 
\begin{equation}
\label{Eq:Hamiltonian_qubit}
    H_0= \frac{\omega_0}{2}(\mathds{1}-\sigma_z) \;, 
\qquad     H_1=\bm{x}\cdot \bm{\sigma}\;,
\end{equation}
 with $\bm{x} := (x_1,x_2,x_3)$ a unit row vector of real components and $\bm{\sigma} := (\sigma_x,\sigma_y,\sigma_z)^T$
 the Pauli column vector.
 
 Let us start by observing  that  whenever $\bm{x}$ is not pointing the $z$-direction,
$H_1$ and $H_0$  form a generator set for the $su(2)$ algebra.
 Accordingly, despite the limited number of charging terms,
 if no restrictions are posed on the intensity of the control function $\lambda_1(t)$ or, in alternative, if the charging time $\tau$ is sufficiently large,  
 the transformations
 (\ref{dfds}) we can induce on the system are still capable of spanning the entire unitary space ${\cal U}$ and one recovers the result~(\ref{anti0}), i.e.
 \begin{eqnarray} \label{unbounded}
E_{\max}(\tau) \Big|_{\text{unbounded}} =  \overline{E}_{\max} = \omega_0\big(\frac{1+|\bm{a}(0)|}{2}\big)\;,
 \end{eqnarray} 
 where, given $\bm{a}(0)$ the Bloch vector of the initial state $\rho(0)$,
 $({1+|\bm{a}(0)|})/{2}$ is the maximum eigenvalue  of such a state.

\begin{figure}[t!]
\centering
\includegraphics[width = \columnwidth]{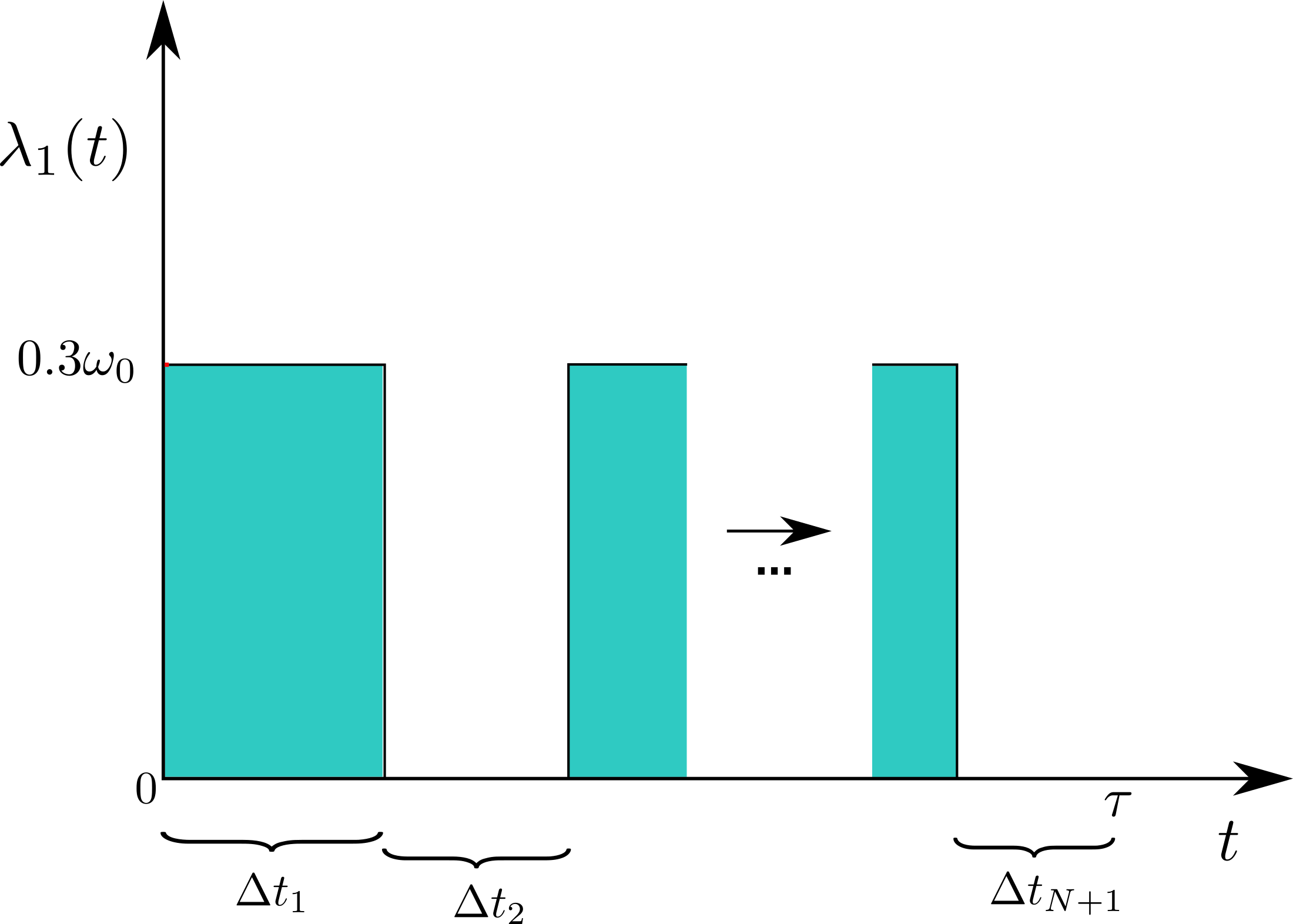}
\caption{Example of an optimal control PMP candidate~(\ref{BANG1}) with 
${N}$ switches for the single-qubit DCP problem with 
$\bm{x}=(1,0,0)$ and $\lambda_1^{\min}=0$: it corresponds to a Bang-Bang function that oscillates between $0$ and $\lambda_1^{\max}$ at the switching times $t_k$ of the selected partition (\ref{partition}).}
\label{figexampprotoc}
\end{figure}
To study the case where instead  $\lambda_1(t)$ is forced to belong to a finite interval~
 ${\cal I}_1= [\lambda_1^{\min}, \lambda_1^{\max}]$, 
   we use the  PMP method detailed at the end of the previous section. 
  In this particular case, $\pi'(\tau)$ is a $2 \times 2$ Hermitian matrix with trace $-\omega_0$. Since the unitary evolution preserves the trace, we can always write the state and the costate as
\begin{equation}
\begin{cases}
\rho(t)=\frac{\mathds{1}+\bm{a}(t)\cdot \bm{\sigma} }{2}\;,\\
\pi'(t)=-\omega_0\frac{\mathds{1}+\bm{b}(t)\cdot \bm{\sigma} }{2}\;,\\
\end{cases}
\label{states}
\end{equation}
where $\bm{a}(t)$ and $\bm{b}(t)$ are two unit row-vectors with  
$\bm{a}(0)$, being the Bloch vector of the input state of the system and $\bm{b}(0)=(0,0,-1)$.
Replacing this into~(\ref{outmean}) and Eq.~(\ref{defGj}) we hence get $E(\tau)= (\omega_0/2)\,(1- {a}_3(\tau)) \;,$
 where ${a}_3(\tau)$ is the $z$ component of $\bm{a}(\tau)$, and 
\begin{equation}
\label{Eq:G0_def} 
\begin{aligned}
G_1(t)=&-i\langle\pi'(t)[H_1,\rho(t)]\rangle=\!\frac{i \omega_0}{4} \langle\bm{b}(t)\!\cdot \! \bm{\sigma}[\bm{x}\cdot \bm{\sigma},\bm{a}(t)\cdot\bm{\sigma}]\rangle\\
=&-\omega_0\;  \bm{b}(t)\cdot \bm{x}\wedge\bm{a}(t) =
\omega_0\;  \bm{x}\cdot \bm{b}(t)\wedge\bm{a}(t) \;.
\end{aligned}
\end{equation}
The crucial case $G_1(t)=0$ can then be translated into the condition 
\begin{equation}
\label{Eq:G0_condition} 
\begin{aligned}
G_1(t)=0 \;\; \Longleftrightarrow \;\;  \bm{x}\cdot \bm{b}(t)\wedge\bm{a}(t)=0\;.
\end{aligned}
\end{equation}
As discussed in App.~\ref{appendix:qubit} this corresponds to the identity
\begin{eqnarray} \label{CONDsingular} 
\lambda^{\star}_1(t)=\frac{\omega_0}{2}x_3\;
\end{eqnarray} 
 as the constraint that leads to a singular interval, with $x_3$ the third component of the unit vector $\bm{x}$.  Following the indications of the PMP detailed in the previous section, we can hence claim that  for the DCP model we are considering here the optimal choice for the control parameter $\lambda^{\star}_1(t)$
 must be indeed a Bang-Bang protocol represented by a piecewise-constant function
 that on the interval $[0,\tau]$  takes values extracted from the three element set  ${\cal S}=\{\lambda_1^{\min}, \, \frac{\omega_0}{2}x_3,\, \lambda_1^{\max}\}$ for $\frac{\omega_0}{2}x_3\in {\cal I}_1$, or  from the two-element set 
${\cal S}=\{\lambda_1^{\min},  \lambda_1^{\max}\}$ if $\frac{\omega_0}{2}x_3\notin {\cal I}_1$.
 Specifically 
giving a $({N}+1)$-elements partition ${\cal P}$ of the charging interval $[0,\tau]$,
\begin{equation} \label{partition} 
0=t_0 <   t_1 < \cdots < t_{{N}} < t_{{N}+1}=\tau\;, 
\end{equation}  
and a collection ${\cal L}:= \{ \Lambda_1, \Lambda_2, \cdots, \Lambda_{{N}+1}\}$ of elements extracted from
the set ${\cal S}$, 
 we can write 
\begin{eqnarray}\label{BANG1} 
\lambda^{\star}_1(t) &=& \sum_{k=1}^{{N}+1} \Lambda_k \; \text{Step}_{\Delta t_k}[t- t_{k-1}] \;, 
\end{eqnarray} 
where for all $k=\{ 1, \cdots, {N}+1\}$  we have
$\Delta t_k:= t_{k}-t_{k-1}$,  and where 
\begin{eqnarray} 
\text{Step}_{\Delta T} [t] &:=& \left\{ \begin{array}{ll}
1 & \forall t\in[0, \Delta T[ \;,\\ \\
0 & \text{otherwise}\;, 
\end{array}
\right. 
\end{eqnarray} 
is a step function of length $\Delta T$. 
 This is clearly a huge simplification of the optimization problem which enlightens the strength of the PMP approach.  Unfortunately the identification of the specific values of 
 ${N}$, ${\cal P}$, and ${\cal L}$ goes beyond the possibility offered by this method and need to be addressed case by case. 
To confirm our theoretical reasoning we performed a numerical simulation for the special case in which 
at at time $t=0$  the battery is in its ground state (i.e. $\rho(0)=\ket{0}\!\bra{0}$ or equivalently
$\bm{a}(0)=(0,0,1)$). Furthermore, to simplify the numerical simulation, we consider the charging Hamiltonian in Eq.\ \eqref{Eq:Hamiltonian_qubit} to be $H_1=\sigma_x$ selecting $\bm{x}=(1,0,0)$, and we fix
$\lambda_1^{\min}=0$ so that the set of allowed pulses ${\cal S}$ reduces to $\{ 0, \lambda_1^{\max}\}$.
With these choices all the candidates for  $\lambda^{\star}(t)$  are given by simple Bang-Bang pulses with alternating
values of  $\lambda_1^{\max}$ and~$0$ (see Fig.~\ref{figexampprotoc}). Excluding the sequences which 
have  $\lambda_1=0$  in the first interval that are clearly sub-optimal (with such a choice nothing is going to happen to the system  at least till $t=t_2$), a complete  parametrization of the PMP candidates (\ref{BANG1}) can hence be obtained in terms of the 
time intervals of the selected partition ${\cal P}$, i.e. 
$\Delta t_1$, $\Delta t_2$, $\cdots$,$\Delta t_{{N}+1}$, such that $\sum_{i=1}^{{N}+1}\Delta t_i=\tau$.
Choosing different sequences of the $\Delta t_k$'s will generate different trajectories and different values of the final energy 
 $E(\tau)=E \big(\Delta t_1,\, \Delta t_2,...,\, \Delta t_{{N}+1}\big)$
 which can be explicitly computed case by case.

\begin{figure*}[htb!]
\centering
\includegraphics[width=2\columnwidth]{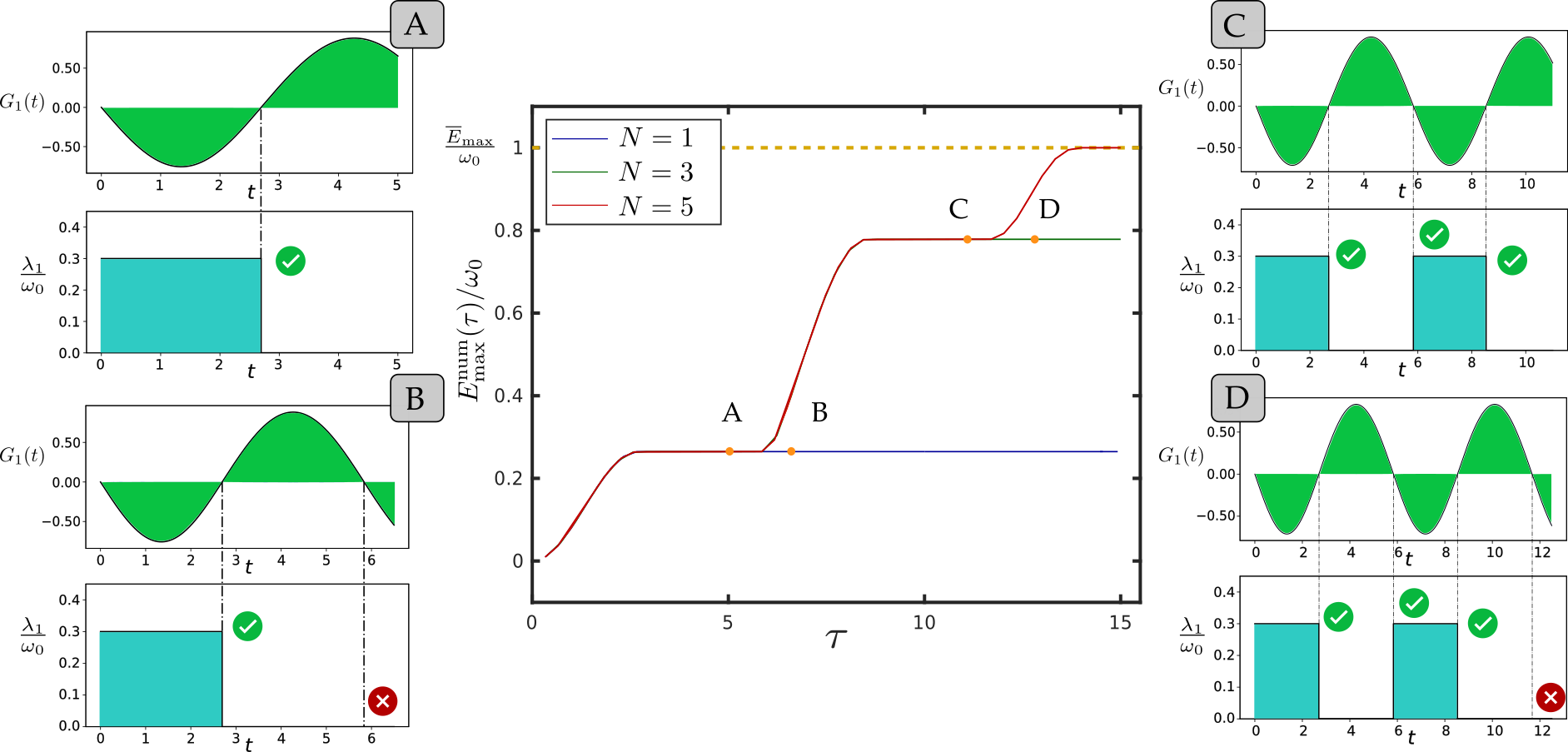}
\caption{Plot of the maximum energy value $E^{\text{num}}_{\max}(\tau)$ at the end of  DCP process~(\ref{Eq:Hamiltonian_qubit})  of duration $\tau$, obtained
by performing a numerical optimization with respect to the Bang-Bang protocols of Fig.~\ref{figexampprotoc}
with different values of ${N}$ (the charging term here is chosen s.t. $\bm{x}=(1,0,0)$). The side panels report also the values of $\lambda_1(t)$ 
 and of the associated $G_1(t)$ function computed as in Eq.~(\ref{Eq:G0_def}) for four particular simulations. 
 Notice that points $A$ and  $C$ follow the  PMP prescriptions detailed at the end of Sec.~\ref{section:onesystem};
 while points $B$ and $D$ do not. More specifically, $B$ and $D$  miss the last switches, continuing to maintain  $\lambda_1(t)=0$: this happens because they have no possible switches left (we have fixed ${N}=1$ and ${N}=3$ respectively,   and, consequently, they are forced to stay with $\lambda_1(t)=0$ for the remaining time).}
\label{fig:simulation}
\end{figure*}

Setting $\lambda_1^{\max} = 0.3 \omega_0$,  we have run a numerical simulation of the problem for different values of the total charging time $\tau$ selected in the domain 
 $[0, 15/\omega_0]$, with 
 different set of time-intervals  $\Delta t_j$'s. 
The obtained results are summarized in 
 Fig.~\ref{fig:simulation} which reports the maximum $E^{\text{num}}_{\max}(\tau)$ of the final energy 
 $E \big(\Delta t_1,\, \Delta t_2,...,\, \Delta t_{{N}+1}\big)$ we have 
 obtained by running a numerical search on Bang-Bang functions of the type Fig.~\ref{figexampprotoc} organized in groups of increasing values of $N$. Specifically, the blue curve 
 reports results obtained  for $N\leq 1$ (i.e.\  piecewise-constant functions with up to 2 intervals $\Delta t_j$'s), the green curve those for $N\leq 3$, and 
  the red curve those with  $N\leq 5$.
The first thing that one can notice is that, as predicted in Eq.~(\ref{unbounded}),
 for $\tau$ sufficiently large (specifically for $\tau \gtrsim  14.0 \omega_0$),
$E^{\text{num}}_{\max}(\tau)$ reaches the value of $\omega_0$, which for the selected choice of the input state corresponds to the absolute maximum $\overline{E}_{\max}$. The plot shows also that
in order to achieve this results we had to use piecewise-constant functions with $N=5$. At the contrary, having $N=1$ or $N=3$ is just enough to push $E^{\text{num}}_{\max}(\tau)$ up to $\sim 0.26 \overline{E}_{\max}$ and $\sim 0.78 \overline{E}_{\max}$ (blue and green plateaus in Fig. \ref{fig:simulation}).
Another element which emerges from the above discussion is that, even though
$E^{\text{num}}_{\max}(\tau)$  is explicitly non decreasing in $\tau$, it exhibits a
staircase-like behaviour with extended plateau regions. 
This means that increasing the final time does not necessarily leads to an increment of the final energy.
 At the contrary, by allowing for negative values of the intensity of the control, one can drastically increase the performance, getting rid of the plateaus and obtaining a monotonically increasing function for $E^{num}_{\max}(\tau)$ (this will be extensively discussed in section \ref{sec:opt2}).
 Naturally, since the analysis relies on a numerical optimization performed on a selected class of Bang-Bang functions with a limited (up to $N+1=6$) number of switching times, one cannot exclude that enlarging the pool of candidates (e.g. increasing $N$) would also remove the staircase behaviour; yet we believe that this is a typical feature of the constraint on the intensity of the control we have chosen, an interpretation that is validated by the fact that $E^{num}_{\max}(\tau)$ is not staircase-like if we allow $\lambda_1(t)$ to be negative.
Our final remark concerns the consistency of the obtained numerical results  with the PMP criteria. For this purpose we focused on  four particular points $A,\,B,\,C,\,D$ of the central plot in Fig.~\ref{fig:simulation}.
Each point corresponds to a particular charging protocol (i.e.\ ${N}$  and $\Delta t_1,\, \Delta t_2,...,\, \Delta t_{{N}+1}$ fixed) for which we present 
the explicit value of $\lambda_1(t)$ and the associated $G_1(t)$ function computed as in Eq.~(\ref{Eq:G0_def}).
We notice that only the protocols on the red line (i.e. $A$ and $C$), which provide our best guess for the maximum final  energy,   fulfil the PMP criteria a), b) c) in Sec.~\ref{PMPenergy}, that prescribe a switch of $\lambda_1(t)$ whenever $G_1(t)$ changes sign. The other two instead fail to follow the  prescription, e.g. missing the final switching point. This is in line with the fact that $B$ and $D$ are clearly not optimal, since there are other points along the red line providing better final energy values for the same charging time $\tau$.

 \subsubsection{Optimal charging times} \label{sec:optimalchargingtime} 
 We now tackle the problem of minimizing the charging time $\tau$ that enables us to reach a  
 final target state $\rho_{\diamond}$. 
Integrating the equation of motions for $\rho(t)$ and $\pi(t)$ given in Sec.~\ref{PMPtime} we expressed them in the
Bloch vector representation, 
\begin{equation}
\begin{cases}
\rho(t)=\frac{\mathds{1}+\bm{a}(t)\cdot \bm{\sigma} }{2}\;,\\\\
\pi(t)= \frac{ b_0 \mathds{1}-\bm{b}(t)\cdot \bm{\sigma} }{2}\;,\\
\end{cases}
\label{states2}
\end{equation}
where at variance with (\ref{states}) we parametrized the  costate in a such a way to leave its trace undetermined
and not directly connected with the length of the vector $\bm{b}(t)$. 
Replacing this into (\ref{defGnuovo}) we hence get 
\begin{equation}
G_1(t)= \frac{i  }{4} \; \langle\bm{b}(t)\cdot\bm{\sigma}[\bm{x}\cdot \bm{\sigma},\bm{a}(t)\cdot\bm{\sigma}]\rangle\\
=  \bm{x}\cdot \bm{b}(t)\wedge\bm{a}(t) \;,\label{Eq:G0_def1} 
\end{equation}
which up to an irrelevant scaling factor $\omega_0$ coincides with the one given in Eq.~(\ref{Eq:G0_def}). 
We can hence apply the same analysis of the previous section to declare that the optimal pulses will be again a
piecewise-constant function belonging to the class~(\ref{BANG1})  with the same set ${\cal S}$ of allowed constant plateaus (see App.~\ref{singularapp} for details).

\subsection{Qubit Optimal DCP with two charging fields (\texorpdfstring{$m=2$}{Lg})} \label{sec:opt2}

We now consider the charging process of a qubit in the presence of two controls. As in Subsec.~\ref{qubit_optimal}, we choose $H_0= \frac{\omega_0}{2}(\mathds{1}-\sigma_z)$ as reference Hamiltonian, but assume the presence of two different charging terms  
 $H_1=\sigma_x$ and $H_2=\sigma_y$ with controls functions  $\lambda_1(t)$ and $\lambda_2(t)$ 
 fulfilling a  constraint which limit their joint intensity, i.e. 
 \begin{eqnarray} 
 \label{eq:constraint_m2}
 \lambda^2_1(t) + \lambda^2_2(t)\leq r^2_{\max}\;,
 \end{eqnarray} 
that we can parametrize as $\lambda_1(t) := r(t)\cos\theta(t)$ and $\lambda_2(t) := r(t)\sin\theta(t)$ with $r(t)\in[0,r_{\max}]$ and $\theta(t)$ real. 
In this case we find it useful to study the problem using the interaction picture where,
given  $V(t) := e^{-iH_0t}$ the unitary associated with the  free evolution, we replace $\rho(t)$ with
the density operator $\tilde{\rho}(t):=V(t)^\dagger \rho(t) V(t)=({\mathds{1}+\tilde{\bm{a}}(t)\cdot \bm{\sigma} })/{2}$, with $\tilde{\bm{a}}(t)$ being its associated Bloch vector. 
Accordingly 
 the dynamical equation 
of the model writes 
\begin{equation} 
\dot{\tilde{\rho}}(t) = -i [\tilde{H}_{\text{INT}}(t), \tilde{\rho}(t)]\quad 
\Longleftrightarrow \quad \dot{\tilde{\bm a}}(t)=2 \tilde{\bm \lambda}(t) \wedge \tilde{\bm a}(t)\;, 
\label{equest} 
\end{equation}
where 
\begin{eqnarray}
	\tilde{H}_{\text{INT}}(t) &:=& V(t)^\dagger \left[ r(t)\cos\theta(t)\sigma_x + r(t)\sin\theta(t)\sigma_y\right]
	  V(t)~, \nonumber \\
	  &=& \tilde{\bm \lambda}(t) \cdot {\bm \sigma}\;,
	\label{eq:h1_tilde}
\end{eqnarray}
is  the interaction picture Hamiltonian characterized by a control vector  $\tilde{\bm \lambda}(t) = (\tilde{\lambda}_1(t),\tilde{\lambda}_2(t), 0)$ of components 
$
\tilde{\lambda}_1(t) := r(t)\cos\tilde{\theta}(t)$,  $\tilde{\lambda}_2(t) := r(t)\sin\tilde{\theta}(t)
$
with $\tilde{\theta}(t) := \theta(t) + \omega_0 t$. 
Noting that 
 the final energy of the system still writes as  \begin{equation} E(\tau)=\langle{\tilde{\rho}(\tau)H_0}\rangle \;, \end{equation}  
we can cast the PMP using an  associated costate $\tilde{\pi}'(t)=
-\omega_0 ({\mathds{1}+\tilde{\bm{b}}(t)\cdot \bm{\sigma} })/{2}$,  which evolves via the same dynamical equation $\tilde{\rho}(t)$, i.e.
\begin{equation} 
\dot{\tilde{\pi}}'(t) = -i [\tilde{H}_{\text{INT}}(t), \tilde{\pi}'(t)]\quad 
\Longleftrightarrow \quad \dot{\tilde{\bm b}}(t)=2 \tilde{\bm \lambda}(t) \wedge \tilde{\bm b}(t)\;,  \label{equest0} 
\end{equation}
and a 2D vector $\tilde{\bm{G}}(t)$ for the corresponding pseudo-Hamiltonian~(\ref{defiH}) that can be expressed 
as 
\begin{eqnarray} 
\tilde{G}_j(t) := \omega_0\; \hat{\bm x}_j \cdot \tilde{\bm{b}}(t)\wedge\tilde{\bm{a}}(t)\;, \qquad \forall j=1,2
\end{eqnarray} 
with $\hat{\bm x}_1=(1,0,0)$ and $\hat{\bm x}_2=(0,1,0)$.
Dropping the irrelevant constant factor $\omega_0$, we can then  cast the third PMP inequality of (\ref{condizionifin}) as 
\begin{eqnarray}\label{condPMP}
 \tilde{\bm{\lambda}}^{\star}(t)  \cdot \tilde{\bm{b}}(t)\wedge\tilde{\bm{a}}(t) \leq  \tilde{{\bm{\lambda}}}(t)  \cdot \tilde{\bm{b}}(t)\wedge\tilde{\bm{a}}(t) \;.
\end{eqnarray} 
 Solving Eq.~(\ref{condPMP}) is much more demanding than the corresponding case with a single control function, so we will adopt a different strategy by guessing the optimal solution and after verifying that it fulfills the necessary conditions \eqref{condPMP}.
Since $E(\tau)=\omega_0(1-\tilde{a}_3(\tau))/2$ we notice that
 increasing $E(\tau)$ is equivalent to decrease the value of $\tilde{a}_3(\tau)$.
 In view of this fact we expect the maximum charging power to be achieved when $\boldsymbol{\lambda}(t)$ is chosen in order to force a rotation of the system (in the interaction picture) around the axis
orthogonal to the plane
containing the $z$-axis and the Bloch vector $\bm{a}(0)$ (note that $\bm{a}(0)$ and $\tilde{\bm{a}}(0)$ coincide).
This axis is 
\begin{eqnarray} \hat{\bm k} :=  \hat{\bm x}_3\wedge \bm{a}(0)/|\bm{a}(0)| = (\cos\theta_0 ,\sin \theta_0,0)\;,\label{defK} 
\end{eqnarray}
with the implicit convention that if $\bm{a}(0)$ is oriented along the $\hat{\bm x}_3$ axis,
we then take $ \hat{\bm k}=(1,0,0)$ (any other  vector orthogonal to  $\hat{\bm x}_3$
would work as well in this case).
To achieve this we need $\tilde{\theta}(t) = \theta_0$ that is realized with the choice
\begin{eqnarray}
\theta(t) =- \omega_0 t + \theta_0 \;, \quad
 r(t) = r_{\max}\;,\label{suitablecontr}
\end{eqnarray} 
where with the second condition we aim at maximizing the speed of rotation.
When \eqref{suitablecontr} holds we have $\tilde{H}_{\text{INT}}(t)=r_{\max} \hat{\bm k} \cdot \bm{\sigma}$ and the dynamical equation simply reads
\begin{eqnarray}  \label{defao} 
&&\tilde{\bm a}(t) = {\bm a}(0) \cos ( 2 r_{\max} t) + ( \hat{\bm k}\wedge {\bm a}(0) ) \sin ( 2 r_{\max} t)\\
&&= \!|{\bm a}(0)| \!\!\left[ \cos( 2 r_{\max} t + \alpha_0) \hat{\bm x}_3\! - \sin(2  r_{\max} t + \alpha_0) 
 ( \hat{\bm x}_3 \wedge \hat{\bm k}  )\right]\!, \nonumber
\end{eqnarray} 
where in the second identity we used~(\ref{defK}) and introduced the symbol  
\begin{equation} \alpha_0 := \arccos\left(\frac{{\bm a}(0) \cdot \hat{\bm x}_3}{|{\bm a}(0)|}\right)=\arccos a_3(0)\in [0,\pi]\;, \end{equation} 
to indicate the angle between the vectors ${\bm a}(0)$ and $\hat{\bm x}_3$.
Accordingly we can write 
$
\tilde{a}_3(t) ={ \bm x}_3\cdot \tilde{\bm a}(t) = |{\bm a}(0)|
\cos(2 r_{\max} t + \alpha_0)$,
so that 
\begin{eqnarray} 
E(\tau)=\omega_0(1-|{\bm a}(0)| \label{evola4} 
\cos(2 r_{\max} t + \alpha_0))/2\;. 
\end{eqnarray} 
Notice now that that for $t$ equal to 
$ \tau_1:=(\pi-\alpha_0)/(2 r_{\max}) $ the function~(\ref{evola4}) reaches its maximum absolute value, i.e.
$\overline{E}_{\max}:=
\omega_0 \tfrac{1+|\bm{a}(0)|}{2}
$.
We can hence  identify two possible scenarios: 
\begin{itemize}
\item  if $\tau\geq  \tau_1$ the optimal protocol is arguably to do a "pi pulse" and keep evolving the system using Eq.~(\ref{suitablecontr}) till $t=\tau_1$, and then stopping, i.e. \begin{eqnarray} 
\tilde{\bm{\lambda}}^{\star}(t) =\left\{ \begin{array}{lll} r_{\max} \hat{\bm k} && \forall t\in[0,\tau_1]\;,\\ \\ 
0 & &\forall t\in]\tau_1,\tau]\;,  \label{solu1} 
\end{array} \right.
\end{eqnarray} 
with an associated  final maximal energy $E(\tau)$ that saturates to the absolute maximum $\overline{E}_{\max}$;
\item 
if $\tau< \tau_1$, our best candidate to the optimal protocol is to  use (\ref{suitablecontr}) till the very end of the charging period, i.e. 
\begin{eqnarray} \label{solu2} 
\tilde{\bm{\lambda}}^{\star}(t) = r_{\max} \hat{\bm k}\;, \qquad  \forall t\in[0,\tau]\;,
\end{eqnarray} 
with an associated optimal final energy that can be estimated as 
\begin{eqnarray}
\label{eq:Emax_m2}
E_{\max}(\tau) = \omega_0\Big[ \frac{1- |{\bm a}(0)|
\cos( 2 r_{\max} \tau + \alpha_0)}{2}\Big]\;. 
\end{eqnarray} 
\end{itemize} 

\begin{figure}[htb!]

  \begin{minipage}[c]{\linewidth}
  \centering
 \begin{overpic}[width=\linewidth]{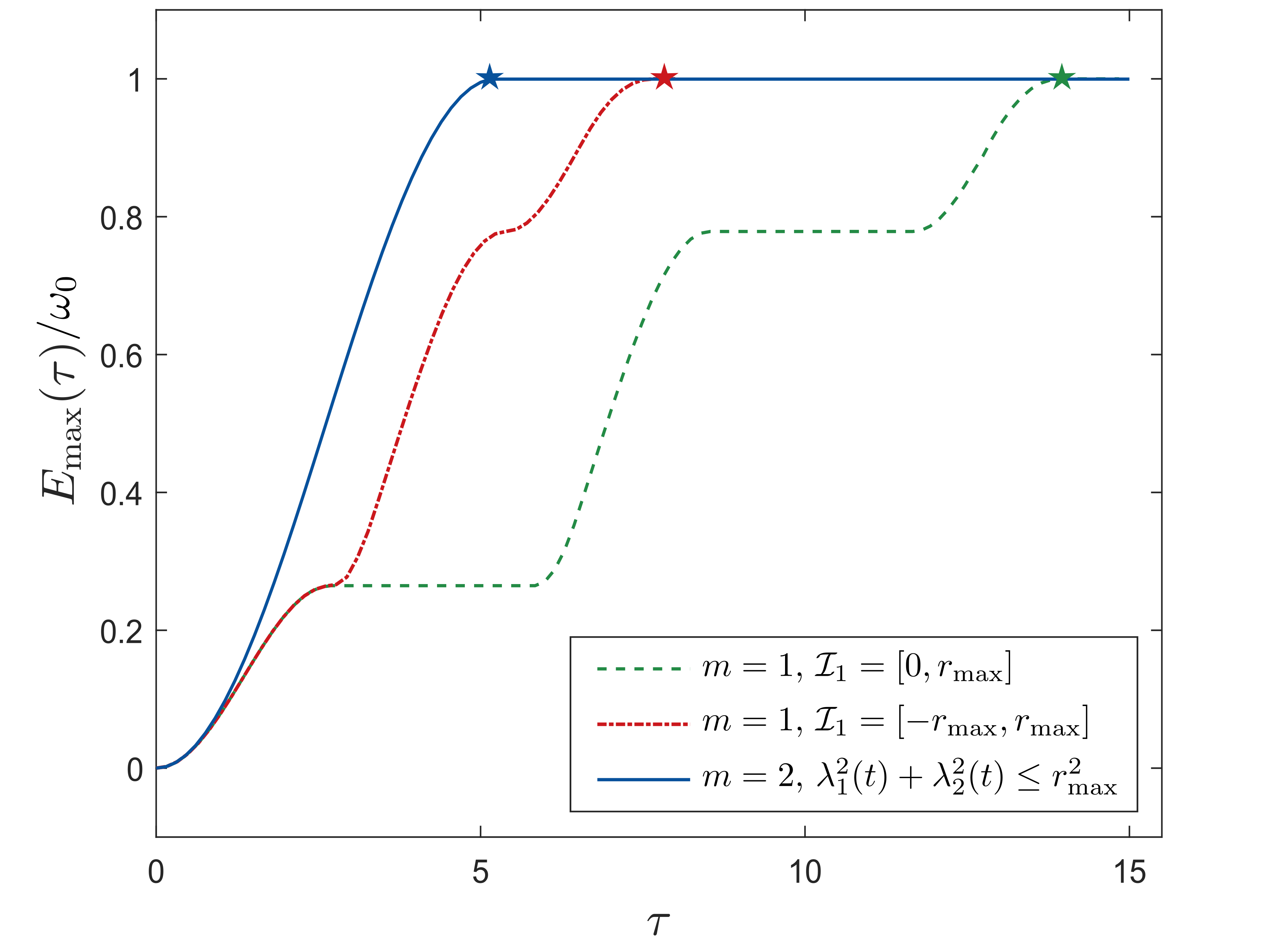}
 \put(1,70){(a)}
 \end{overpic}
  \end{minipage}
  
     \begin{minipage}[c]{\linewidth}
  \centering
 \begin{overpic}[width=0.95\linewidth]{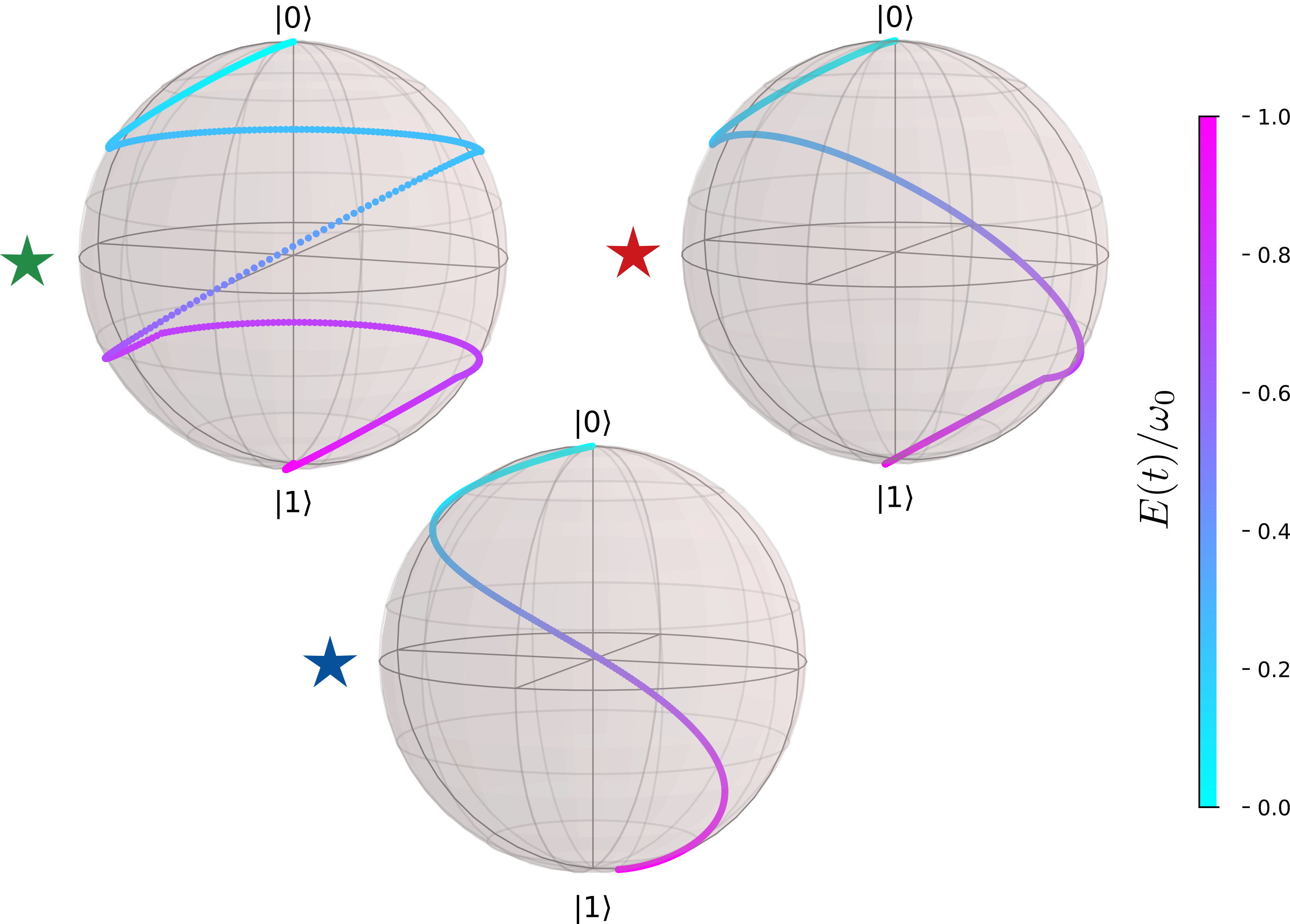}
 \put(1,70){(b)}
 \end{overpic}
  \end{minipage} 
  \caption{Comparison of optimal charging processes for three different DCP models corresponding to different constraints on the controls (see legend). Panel (a) shows the maximum energy value at the end of a charging process of duration $\tau$, where we have set $r_{\max}= 0.3 \omega_0$. Panel (b) shows the evolution in the Bloch's sphere of a full charging process for all the different DCP models. In particular, each Bloch's sphere represents a specific instance of the optimal charging processes displayed in panel (a) (represented by a green, red and blue star).}
  \label{fig:DCP_comparisons}
\end{figure}
We finally checked (see \ref{newapp} for the details) that the above guesses verify the constraint Eq.~(\ref{condPMP}). 
To conclude, it is interesting to compare the energy achieved with the optimal protocol in the $m=2$ with its analog in the $m=1$ case (discussed in sec. \ref{qubit_optimal}), thus highlighting the advantage of an increased accessible domain for the charging hamiltonians.
In Fig.\ \ref{fig:DCP_comparisons} we plot the final energy in Eq.\  \eqref{eq:Emax_m2}, picking $r_{\max}= 0.3 \omega_0$, for different values of the total charging time $\tau$ and initializing the system in the ground state. In addition, we plot the correspondent quantity $E_{max}(\tau)$ related to the case of only one control field ($m=1$) treated in Sec. \ref{qubit_optimal}  and an alternative $m=1$ case in which we allow the intensity of the control to be negative. As expected, the latter two cases are sub-performing with respect to the $m=2$ case.

\subsection{Harmonic Oscillator Optimal Charging} \label{sec:harmoniccharge}
Here we analyze  a DCP model for continuous variable system with a single excitation mode described by the usual Hamiltonian \eqref{eq:charging_model}, with a single control function ($m=1$) and 
\begin{equation}
H_0= \omega_0 a^{\dag} a \;, \qquad 
H_1= a + a^{\dag} \;, 
\label{hamosc}
\end{equation}
 where $a^{\dagger}$ ($a$) is the creation (destruction) bosonic operator.
As for the qubit DCP model of Sec.~\ref{qubit_optimal}, we are interested in finding the optimal function $\lambda_1^{\star}(t)$, with $\lambda_1^{\min} \le \lambda_1(t) \le \lambda_1^{\max}$ as a constraint, that enforces an evolution that maximizes the energy of the system in a fixed time $\tau$. 
In this  case instead of solving the dynamical evolution in the standard Schr\"{o}dinger 
picture
we find it useful to adopt 
the Heisenberg representation.
The reason for such a choice is that 
the expectation values of the first and second momenta of the field operator form 
a closed system of differential equations, i.e. 
\begin{equation}
\begin{array}{l}
v_1(t) =\langle a^{\dag}a\rho(t) \rangle \\
v_2(t) =Im\, \langle a\rho(t) \rangle \\
v_3(t) =Re\,\langle a \rho(t) \rangle
\end{array}
\,\implies \,
\begin{cases}
\dot{v}_1(t)=-2\lambda_1(t)v_2(t)\;,\\
\dot{v}_2(t)=-\omega_0 v_3(t)\! -\! \lambda_1(t)\;,\\
\dot{v}_3(t)= \omega_0 v_2(t)\;,
\end{cases}
\label{motostati}
\end{equation}
with the cost function~(\ref{THECOST}) expressed as
\begin{equation}
J=-\omega_0\int^{\tau}_0 \dot{v}_1(t)\, dt\;.
\label{eq:e_harmonic}
\end{equation}
Notice that Eq.~(\ref{motostati}) represents the equation of motion of a classical harmonic oscillator driven by an external time-dependent force proportional to $\lambda_1(t)$, and Eq.~(\ref{eq:e_harmonic}) is proportional to the energy of the classical oscillator. Indeed, denoting with $r(t)$ and $v(t)$ the position and velocity of a particle of mass $m$ coupled to a spring characterized by $k=m\omega_0^2$, the second equation of~(\ref{motostati}) can be written as $m\dot{v}(t) = -kr(t) + F(t)$ through the identification
   $v_2(t) = \omega_0^{-1} v(t)$, 
    $v_3(t) = r(t)$, $\lambda_1(t) = -(m\omega_0)^{-1} F(t)$, while the third equation on  
 the power of the battery being proportional to the power of the classical harmonic oscillator, i.e. $\dot{E}(t) = 2(m\omega_0)^{-1}v(t)F(t)$.

We now turn to the optimal control setting.
We define the pseudo-Hamiltonian as in Eq.\ \eqref{pseudohamgen}:
\begin{equation}
\begin{aligned}
\mathcal{H}= & 2\omega_0 \lambda_1(t) v_2(t) + p_1(t)[-2 \lambda_1(t) v_2(t)]\\
&+ p_2(t)[-\omega_0 v_3(t) - \lambda_1(t)] + p_3(t)[\omega_0 v_2(t)]\;,
\end{aligned}
\end{equation}
where the $p_i(t)$'s are the costates that enforce the evolution of the first and second momenta.
Notice that as we are still in the case where $\mathcal{H}$ is linear in the control
function $\lambda_1(t)$, the PMP inequality~(\ref{cond4})  is still of the form~(\ref{condizionifin})  
\begin{eqnarray}
{{\lambda}}_1^{\star}(t) {{G}}_1(t) \leq  {{\lambda}}_1(t) {{G}}_1(t)\;, 
\end{eqnarray} 
with a single function 
\begin{equation}\label{defg1} 
G_1(t)= 2\omega_0v_2 (t)-2p_1(t)v_2(t)-p_2(t)\;.
\end{equation}
As proved in the Appendix \ref{appendix:harmonic_oscillator}, when starting from the ground state of $H_0$, no singular intervals are allowed, leading to an optimal charging protocol consisting of a Bang-Bang modulation with 
$\lambda_1(t)$ switching between the values  $\lambda_1^{\min}$ and $\lambda_1^{\max}$.
In the long time limit, this modulation achieves the optimal performance when its frequency is resonant with the one of the oscillator, as proven in Appendix \ref{app:oscopt}.

\section{Examples of MCP models}
\label{section:charger-mediated}
In this section we analyze MCPs where energy is transferred to the battery through an additional system (charger).

\subsection{Qubit - Qubit}
We begin by studying the most straightforward case of a charger-battery setting with a single controllable interaction term ($m=1$): here, the charger and the quantum battery are two qubits that evolve according to a global Hamiltonian  of the form \eqref{eq:nuovach}, with
 \begin{equation}
 \begin{array}{l}
 H_A= \frac{\omega_A}{2}(\mathds{1}-\sigma_z^{\rm A})\;,\\[0.1cm]
 H_B= \frac{\omega_B}{2}(\mathds{1}-\sigma_z^{\rm B})\;,\\[0.1cm]
 H_1=(\sigma^{\rm A}_+ + \sigma^{\rm A}_-)(\sigma^{\rm B}_++\sigma^{\rm B}_-)\;,\\
 \end{array}
 \label{hamdoubleelenco}
 \end{equation}
where $\sigma^{\rm S}_{x,y,z}$ are Pauli matrices acting on system $S=A,B$ and $\sigma^{\rm S}_+= [\sigma^{\rm S}_-]^{\dagger} = (\sigma^{\rm S}_x+ i \sigma^{\rm S}_y)/2$ is the  two-level raising operator.
Throughout this section, we  focus on the  $\omega_A \neq \omega_B$ case since the energy transfer trivially occurs via the well-known Rabi oscillations \cite{Qbattery1} 
when the two qubits are resonant ($\omega_A= \omega_B$). 
 For  general  initial states, determining the optimal $\lambda_1(t)$ that leads to the maximum value for the final
 energy stored in the subsystem $B$
 is quite challenging, due to the fact that the evolution of the quantum battery in this setting is not unitary. However, we can exploit the fact that the in the $\big\{\ket{00},\ket{11},\ket{10},\ket{01}\big\}$ basis the resulting Hamiltonian \eqref{Hamdouble} is a block diagonal matrix:
 \begin{equation}
H(t)=
\begin{pmatrix}
   0 & \lambda_1 (t) & 0 & 0 \\
   \lambda_1 (t) & \omega_B+\omega_A & 0 & 0\\
   0 & 0 & \omega_A& \lambda_1 (t)\\
   0 & 0 & \lambda_1 (t) & \omega_B
\end{pmatrix}\;.
\label{Hamdouble}
\end{equation}
Accordingly, we can map the MCP model into a single-qubit DCP scheme by suitably choosing the initial state.

\subsubsection{Case \texorpdfstring{$\rho_{AB}(0)=\ket{10}\!\bra{10}$}{Lg}}
\label{section:charger_full}
We first consider the  battery in the ground state and the charger  completely charged, assuming
as input state of the model $\rho_{AB}(0)=\ket{10}\!\bra{10}$. 
It is evident that in this situation we can consider just the second block in Eq.~\eqref{Hamdouble}, associated with the basis $\{\ket{10},\,\ket{01}\}$.
Let us call our new vector basis as $\ket{g}:= \ket{10}$ (for "ground state") and $\ket{e} := \ket{01}$ (for "excited state").
This is now equivalent to a single-qubit model with reference Hamiltonian 
\begin{equation}
\label{refhamiltonian}
H'(t)  :=H'_0+ \lambda_1(t) H_1'\;,
\end{equation} 
 where
\begin{equation}
\begin{array}{l}
H'_0:= \frac{\omega_A+\omega_B}{2}\mathds{1} + \frac{\omega_A-\omega_B}{2}\sigma_z
= \omega_A |g\rangle\langle g| + \omega_B |e\rangle\langle e| \;, \\
H'_1:= \sigma_x = |e\rangle\langle g| + |g\rangle\langle e| \;.
\end{array} 
\label{hampsin10}
\end{equation}
 The global state at time $t$ can be written as $\ket{\psi'(t}) = \alpha(t)\ket{g} + \beta (t) \ket{e}$, 
corresponding to a reduced density matrix   $\rho_B(t) = 
|\alpha(t)|^2 \ket{0}\!\bra{0}
 + |\beta (t)|^2 \ket{1}\!\bra{1}$ for the battery.
The maximization of $E_B(\tau)$ can now be turned into a  DCP problem by noting that 
 \begin{eqnarray}
 E_B(\tau)&=&\langle\rho_B(\tau) H_B \rangle=|\beta (\tau)|^2 \omega_B \nonumber \\
 &=& \langle\rho'(t) H'_B \rangle \;,
 \end{eqnarray}
 with $H'_B:=\omega_B \ket{e}\!\bra{e}$. 
The original MCP has been turned into a modified single-qubit DCP problem, that is the same as the one presented in section \eqref{qubit_optimal}, apart from an additional term appearing in the energy function $H_B'=H_0'- H_A'$, where $H_A'= \omega_A \ket{g}\bra{g}$.
However, notice that the presence of $H_A'$ does not change the nature of the optimal solutions, since the points where $E_B(\tau) = |\beta(\tau)|^2 \omega_B$ and $\langle \rho'(t) H_0' \rangle = \omega_A + |\beta(\tau)|^2 (\omega_B-\omega_A)$ have extrema are the same (more in detail, a maximum of the former respectively corresponds to a maximum or minimum of the latter depending on the sign of $\omega_B-\omega_A$).
With this in mind, we can still treat the problem using the same PMP approach we detailed in Sec.~\ref{section:onesystem}, writing the cost function as
\begin{equation}
J=-\int^{\tau}_0 \langle H'_B \mathcal{N}[\rho'(t)]\rangle dt\;,
\end{equation}
with $\mathcal{N}[\rho'(t)]= -i\big[ H'(t), \rho'(t)\big]$. 
 More precisely, the optimization problem is equivalent, with the following arrangements:
 \begin{align}
& \ket{0} \rightarrow \ket{g}\;, \quad \quad 
 \ket{1} \rightarrow \ket{e}\;, \\
& H_0 \rightarrow H_0'\;, \quad \quad 
 H_1 \rightarrow H_1'\;,\\
&  E(t)=\langle\rho(t)H_0\rangle \rightarrow E_B(t)=\langle\rho'(t) H'_B \rangle\;.
\end{align}
Therefore, since $H_1' = \sigma_x$,  we have shown in section \ref{qubit_optimal} that Bang-Bang-off protocol is the optimal choice for this initial configuration.

\subsubsection{Case \texorpdfstring{$\rho_{AB}(0)=\ket{00}\!\bra{00}$}{Lg}}
We now consider a case where both the battery and the charger start in their ground state. The process is not a simple energy flow from one system to another since they start completely uncharged. Instead, the energy comes from the modulation of the interacting Hamiltonian $H_1$. The charger works more like a  ``plug" that allows the battery to absorb energy thanks to their interaction.
Interestingly, thanks to the block structure of the global Hamiltonian \eqref{Hamdouble}, also this case can be mathematically mapped into a single-qubit battery with reference Hamiltonian as in \eqref{refhamiltonian}, where
\begin{equation}
\begin{array}{l}
H'_0= \frac{\omega_A+\omega_B}{2}\mathds{1} - \frac{\omega_B+\omega_A}{2}\sigma_z\;,\\
H'_1= \sigma_x\;.
\end{array}
\end{equation}
Therefore, the optimal charging  with these initial conditions will still be performed through a Bang-Bang-off protocol.

\subsection{Harmonic Oscillator - Qubit}

In this section we want to understand if it is possible to boost the qubit charging process by considering a different charger, focusing on the analysis of a quantum harmonic oscillator system as a charger. Since we can have more than one excited level in this case, we expect that fixing the same frequency $\omega_A$ will allow us to charge our battery faster. We consider the global Hamiltonian of the system  still in the form  \eqref{eq:nuovach}, with:
\begin{gather}
 H_A=\omega_A \, a^{\dag}a,  \quad \quad H_B = \omega_B \frac{\mathds{1}-\sigma_z}{2}\;, \\[0.1cm]
H_1=a^{\dag}\sigma_- + a \sigma_+\;.
\end{gather}

Considering as initial state  $\rho_{AB}(0)=\ket{n,0}\!\bra{n,0}$ that is, the charger is prepared in an eigenvector of the number operator $a^{\dag} a$ with eigenvalue $n$, and the battery is initialized in the ground state,
we can restrict the analysis to subspaces with a given number $n$ of excitations spanned by vectors $\ket{g}=\ket{n,0}$ and $\ket{e}=\ket{n-1,1}$. The Hamiltonian contributions  in this two-dimensional subspace are
 \begin{equation}
 \begin{array}{l}
 H_0'=\big[\frac{\omega_A(2n-1)+\omega_B}{2}\big] \mathds{1} + \big(\frac{\omega_A-\omega_B}{2}\big) \sigma_z\;,\\
H_1'=\sqrt{n}\sigma_x\;.
 \end{array}
 \end{equation}
 This is equivalent  to the  qubit-qubit case in Eq.\ \eqref{hampsin10} by changing $\lambda^{\max} \, \rightarrow \, \sqrt{n}\lambda^{\max}$ (the coefficient that multiplies the identity operator is always irrelevant).
 Consequently, we are boosting the Bang-Bang-off protocol, allowing to charge the qubit battery faster than the two-qubit's protocol by a factor of $\sqrt{n}$.

\section{Conclusions}
\label{section:conclusion}

We have presented a systematic analysis of two quantum battery charging processes, focusing on qubit systems and quantum harmonic oscillators. We analyzed two charging scenarios.
\begin{itemize}
\item Direct charging process, where a single quantum system, representing the battery, is charged through the modulation of an external Hamiltonian.
\item Mediated charging process, where energy is transferred between two quantum systems A and B,  representing respectively the charger and battery. 
\end{itemize}
We have shown that the optimal charging protocols for both approaches are obtained by modulating  the control parameter as a step function between few specific values, greatly simplifying the optimal control problem. In particular, we observed that alternating the intensity of the control parameter between its boundary limits is almost always an optimal strategy.

We have also shown that replacing the qubit charger with a quantum harmonic oscillator can enhance the performance of our charging process, allowing us to charge the battery faster. This result was expected since we can store  more energy in a quantum harmonic oscillator system with the same frequency. This inevitably has a positive impact on the charging protocol, as encountered in our analysis.

A natural direction for future research is extending this analysis to the case of open quantum systems, where a unitary operation no more describes the state evolution. In addition, it would be worth attempting to consider entangled initial states in the charger-mediated process, hoping to provide an additional speed up to the charging process.

\acknowledgments
F.M. acknowledges support by the European Union’s Quantum Technology Flagship, through the Horizon 2020 research and innovation programme, under Grant agreements CIVIQ No 820466 and OPENQKD No 857156. V.C. is supported by the Luxembourg national research fund in the frame of Project QUTHERM C18/MS/12704391. 
P.A.E. gratefully acknowledges funding by the Berlin Mathematics Center MATH+ (AA1-6).
V.G. acknowledges financial support by MIUR (Ministero dell’ Istruzione, dell’ Universit\'a e della Ricerca) by PRIN 2017 Taming complexity via Quantum Strategies: a Hybrid Integrated Photonic approach (QUSHIP) Id. 2017SRN- BRK, and via project PRO3 Quantum Pathfinder.

\appendix
\
\section{Ergotropy, total ergotropy and thermal free-energy}\label{app:defergo} 
At variance with purely classical settings, 
 discriminating which part of the internal energy 
of a quantum system $\rho$ can be identified with extractable work is 
difficult~\cite{Scovil1959,Alicki1979,Kosloff1984,Niedenzu2019}.
Ergotropy ${\cal E}[\rho,H]$, total ergotropy ${\cal E}_{tot}[\rho,H]$ and thermal free-energy 
${\cal F}_{\bar{\beta}}[\rho,H]$ are three different ways to evaluate such quantity 
based on different assumptions on the resources we have dedicated to the task. 
The first one measures the amount of work we can
get from $\rho$ if we limit the allowed operations to local unitary transformations.
Formally it can be expressed as
\begin{eqnarray} \label{def:ergo}
		 {\cal E}[\rho,H] &:=& \langle \rho H\rangle - \min_{U\in {\cal U}} \langle U \rho U^\dag H\rangle \;, 
		\end{eqnarray}
		where the minimization in the first term is performed over all possible unitary transformations acting on the system. Such term can be cast in a closed formula by introducing the passive counterpart $\rho^{\downarrow}$ of $\rho$ \cite{Pusz1978, Lenard1978}, i.e. the special state which has the lowest energy among those with the same spectrum of~$\rho$. Introducing the spectral decomposition $\rho=\sum_i \eta_i |i\rangle\langle i|$ and $H=
		\sum_i \epsilon_i |\epsilon_i\rangle\langle\epsilon_i|$ of the state and of the Hamiltonian, 
we can write 		
	\begin{eqnarray} 
	\rho^{\downarrow} := \sum_{i} \eta^{\downarrow}_i  |  \epsilon^{\uparrow}_i \rangle\langle \epsilon^{\uparrow}_i |
\;,
	\end{eqnarray} 
	where 
	 $s^{\downarrow}_{\rho}:=\{ \eta^{\downarrow}_1, \eta^{\downarrow}_2, \cdots\}$ is a rearrangement of the 
	 spectrum $s_{\rho}:=\{ \eta_1, \eta_2, \cdots\}$
	 of $\rho$ where the various terms are organized in the decreasing order (i.e. $\eta^{\downarrow}_{i} \geq \eta^{\downarrow}_i$), and $\{ |  \epsilon^{\uparrow}_i \rangle\}_i$ are instead the eigenvectors of the system Hamiltonian organized in 
	 increasing order of their associated eigenvalues (i.e. $\epsilon^{\uparrow}_i \leq \epsilon^{\uparrow}_{i+1}$). 
	 With this choice Eq.~(\ref{def:ergo}) can hence be written as 
	 \begin{equation} \label{def:ergo1}
		 {\cal E}[\rho,H] = 	\langle \rho H\rangle - \langle \rho^{\downarrow} H\rangle 
		 = \langle \rho H\rangle  -  \sum_{i} \eta^{\downarrow}_i \; \epsilon_i^{\uparrow}\;, 
		\end{equation}
		which applied to the our problem leads to Eq.~(\ref{dfdf}) with 
		${\cal F}(s_{\rho(t)},s_{H_t})={\cal F}(s_{\rho(0)},s_{H_0})= \sum_{i} \eta^{\downarrow}_i(0)\; \epsilon_i^{\uparrow}(0)$.
		It is worth noticing that a quantity that is related to ${\cal E}[\rho,H]$ is the anti-ergotropy functional ${\cal A}[\rho,H]$ that instead gauges the minimum work extractable from the system via unitary transformations \cite{Salvia2021}. This is obtained by replacing the minimization in Eq.~(\ref{def:ergo}) with a maximization, i.e. 
		\begin{eqnarray} \nonumber 
		{\cal A}[\rho,H] &:=&\langle \rho H\rangle - \max_{U\in {\cal U}} \langle U \rho U^\dag H\rangle\\
		&=&\langle \rho H\rangle - \langle \rho^{\uparrow} H\rangle 
		 = \langle \rho H\rangle  -  \sum_{i} \eta^{\uparrow}_i \; \epsilon_i^{\uparrow}\;, 
		\end{eqnarray} 
		where now $\rho^{\uparrow}:= \sum_{i} \eta^{\uparrow}_i  |  \epsilon^{\uparrow}_i \rangle\langle \epsilon^{\uparrow}_i |$ is the anti-passive counterpart of the $\rho$.

Ergotropy  turns out to be non-extensive~\cite{skrzypczyk2015, Salvia2020}: when operating with reversible coherent operations on  $N$ copies of a given state $\rho$, 
 it is possible  to increase the total amount of extractable energy by acting jointly on the whole set of subsystems.
The maximum amount of energy per copy that is attainable under this new paradigm is quantified by the {\it total ergotropy} ${\cal E}_{tot}[\rho,H]$, a functional  which can obtained via a proper  regularization of~(\ref{def:ergo}), i.e. 
	\begin{eqnarray}
	\label{ergo_tot}
{\cal E}_{tot}[\rho,H] &:=& \lim_{n \to \infty} \frac{1}{n} {\cal E}[\rho^{\otimes n}, H^{(n)}] = 
\langle \rho H\rangle  - \langle \tau_{\beta}  H\rangle 
\nonumber \\
&=& \langle \rho H\rangle  - \frac{\sum_i e^{-\beta \epsilon_i} \epsilon_i}{\sum_i   e^{-\beta \epsilon_i}}\;,
\end{eqnarray}
	where $\tau_{\beta}
 := {e^{-\beta H}}/{\mbox{Tr}[e^{-\beta H}]}$
 is a thermal Gibbs state of the system  whose inverse temperature $\beta \in \mathbb{R}^+$  
	 is fixed in order to ensure  that it posses the same von Neumann entropy of $\rho$, i.e. 
\begin{equation}
S(\tau_{\beta})=S(\rho) :=-\mbox{Tr}[\rho \log \rho] = -\sum_i \eta_i \log \eta_i\;. 
\end{equation}
Notice that as $\beta$ is an implicit function of just the spectrum of $\rho$, we can again cast
the total ergotropy as in Eq.~(\ref{dfdf}) by setting  
${\cal F}(s_{\rho(t)},s_{H_t}) = {\cal F}(s_{\rho(0)},s_{H_0})= \langle \tau_{\beta_0}  H_0\rangle$
with $\beta_0 =\beta(s_{\rho(0)})$. 

Finally beyond the value defined by ${\cal E}_{tot}[\rho,H]$
 more energy from the system can still be converted into useful work only if we are willing to admit
 some dissipation side-effect, e.g. by coupling the system with an external thermal bath~\cite{Esposito2011,Niedenzu2019}. In this case the overall amount of extractable energy is provided by the   
 non-equilibrium {free energy} functional:
	\begin{equation}
{\cal F}_{\bar{\beta}}[\rho,H] := \langle \rho  H \rangle- S(\rho)/\bar{\beta} =
 \langle \rho  H \rangle +(1/\bar{\beta}) \sum_i \eta_i\log \eta_i\;, 
\end{equation}
	with $\bar{\beta}$ representing the inverse temperature of the bath.
	Once more, for our problem the above expression reduces to form 
	Eq.~(\ref{dfdf}) taking		${\cal F}(s_{\rho(0)},s_{H_0})= (1/\bar{\beta})\sum_i \eta_i(0)\log \eta_i(0)$.

\section{Singular Intervals analysis for the DCP qubit model with a single control function}
\label{appendix:qubit}

 We have shown in Sec.~\ref{qubit_optimal} that  for the DCP qubit  model with a single control function 
   having  singular intervals ($G_1(t)=0$)  
   is equivalent to have $\bm{x}\cdot \bm{b}(t)\wedge\bm{a}(t)=0$,
   see Eq.~(\ref{Eq:G0_condition}). A closed inspection of this formula implies that there are only two alternatives allowed:
\begin{itemize}
\item[\bf 1)] $\bm{x} \perp \bm{a}(t)\wedge \bm{b}(t)\;,$
\item[\bf 2)] $\bm{a}(t) \parallel \bm{b}(t)\;,$
\end{itemize} 
with $\bm{x}$ the 3-D vector which define the charging Hamiltonian (\ref{Eq:Hamiltonian_qubit}), and with 
$\bm{a}(t)$ and $\bm{b}(t)$ the Bloch vectors~(\ref{states}) which define the temporal evolution of the state and of the costate of the system~(\ref{Eq:Hamiltonian_qubit}).
In the following we shall analyze separately the two cases showing that the only possible option one has is
provided by the condition~(\ref{CONDsingular}) of the main text.

\subsection*{Condition  {\bf 1)}} 
Enforcing the condition {\bf 1)} for some non trivial temporal interval, requires that in such interval $\bm{a}(t)$ and $\bm{b}(t)$ remain in the the plane orthogonal to the vector $\bm{x}$.  Rewriting the system Hamiltonian in the Bloch vector form,
\begin{eqnarray}
H(t) = H_0 +\lambda_1(t) H_1 = {\bm n}(t) \cdot \bm{\sigma}\;, 
\end{eqnarray} 
with ${\bm n}(t)$ the row vector 
\begin{equation}
{\bm n}(t):= 
          2 (x_1\lambda_1(t), 
           x_2\lambda_1(t) ,
           x_3\lambda_1(t) -\omega_0 /2)          \;, 
      \label{defne}
\end{equation} 
reveals that the dynamics (\ref{condizionifin}) forces both $\bm{a}(t)$ and $\bm{b}(t)$  to undergo to rotations around  the time-dependent axis (\ref{defne}) evaluated on the optimal control  $\lambda_1^\star(t)$, i.e. 
\begin{eqnarray}\nonumber 
\dot{\bm{a}}(t) &=& \langle \bm{\sigma} \dot{\rho}(t)\rangle =-i  \left\langle  \bm{\sigma}\left[ {\bm n}^\star(t) \cdot \bm{\sigma},\rho(t)  \right] \right\rangle = {\bm n}^\star(t) \wedge\bm{a}(t) \;, \\
\dot{\bm{b}}(t) &=&-\tfrac{1}{\omega_0} \frac{\langle \bm{\sigma}\dot{\pi'}(t)\rangle}{\omega_0} =\tfrac{i}{\omega_0}  \left\langle  \bm{\sigma}\left[ {\bm n}^\star(t) \cdot \bm{\sigma},\pi'(t)  \right] \right\rangle \nonumber \\
 &=& {\bm n}^\star(t) \wedge\bm{b}(t) \;,  \label{ff11} 
\end{eqnarray} 
with 
\begin{eqnarray}
{\bm n}^\star(t):={\bm n}(t)\Big|_{\lambda_1(t)=\lambda_1^\star(t)}.
\end{eqnarray}  
A little algebra now reveals that the condition {\bf 1)} allows only one possible solutions i.e. taking $\bm{x}$ orthogonal to ${\bm n}^\star(t)$. To see this explicitly observe that 
by construction Eq.~(\ref{ff11}) implies 
 that also  the vectors ${\bm c}(t) =\bm{a}(t)\wedge \bm{b}(t)$ and all its time-derivative   undergo to the same dynamics of $\bm{a}(t)$ and $\bm{b}(t)$, i.e. 
\begin{eqnarray}\nonumber 
\dot{{\bm c}}(t) &=&{\bm n}^\star(t) \wedge{\bm c}(t) \;, \\
 \ddot{{\bm c}}(t) &=&{\bm n}^\star(t) \wedge\dot{{\bm c}}(t) = {\bm n}^\star(t) \wedge ({\bm n}^\star(t) \wedge{\bm c}(t))\;.
\end{eqnarray} 
Now if we wish to enforce the orthogonality condition between ${\bm c}(t)$ and $\bm{x}$ for some finite time interval
that implies in particular that the following conditions must hold:
\begin{eqnarray}  \label{prima}
\bm{x} \cdot {\bm c}(t) &=& 0\;, \\
\bm{x} \cdot \dot{{\bm c}}(t) &=& \bm{x} \cdot {\bm n}^\star(t) \wedge{\bm c}(t)=0  \;, \\
\bm{x} \cdot \ddot{{\bm c}}(t) &=& \bm{x} \cdot {\bm n}^\star(t) \wedge ({\bm n}^\star(t) \wedge{\bm c}(t)) =0\;, 
\label{terza} 
\end{eqnarray} 
i.e we need to choose $\bm{x}$ in such a way that it is orthogonal to ${\bm c}(t)$, ${\bm c}_1(t):={\bm n}^\star(t) \wedge{\bm c}(t)$
and ${\bm c}_2(t):={\bm n}^\star(t) \wedge ({\bm n}^\star(t) \wedge{\bm c}(t))$ at the same time. 
Since all these vectors live on a 3D space, the only possibility we have to fullfil such constraint is when 
 ${\bm c}(t)$, ${\bm c}_1(t)$, and ${\bm c}_2(t)$ are not linearly indepedent.
 Consider first the scenario where ${\bm n}^\star(t)$ is parallel to ${\bm c}(t)$: in this case
 ${\bm c}_1(t)={\bm c}_2(t)=0$ and the last two conditions of (\ref{prima})-(\ref{terza}) trivialize.
A solution of the problem {\bf 1)} can be hence obtained by forcing orthogonality between $\bm{x}$ and ${\bm n}^\star(t)$, i.e.
\begin{equation}
\bm{x}\cdot {\bm n}^{\star}(t)=
{2 \lambda^{\star}_1(t) - \omega_0 x_3}=0 \;, 
\end{equation} 
leading to the condition~(\ref{CONDsingular}) of the main text. 
 Consider next the case where instead 
  ${\bm n}^\star(t)$ is orthogonal to ${\bm c}(t)$: in this case
 ${\bm n}^\star(t)$, ${\bm c}(t)$, and 
 ${\bm c}_1(t)$ will form an orthogonal set, forcing ${\bm c}_2(t)$   to be parallel to ${\bm c}(t)$. 
 In other words if ${\bm n}^\star(t)$ is orthogonal to ${\bm c}(t)$, then ${\bm c}(t)$, ${\bm c}_1(t)$, and ${\bm c}_2(t)$
 lay on a plane which is orthogonal to ${\bm n}^\star(t)$ and one could satisfy the conditions (\ref{prima})-(\ref{terza})
 by simply choosing $\bm{x}$ parallel to ${\bm n}^\star(t)$. However as evident from (\ref{defne}) the only case where we can have $\bm{x} \parallel {\bm n}^\star(t)$, is when $\omega_0=0$, which is not included in our analysis. 
Finally we are left with the intermediate case where ${\bm n}^\star(t)$ is neither orthogonal nor parallel to ${\bm c}(t)$: in this scenario we shall have that ${\bm c}(t)$, ${\bm c}_1(t)$, and ${\bm n}^\star(t)$ will be independent but
will not form a mutually orthogonal set. Therefore in this case ${\bm c}_2(t)$ is not forced to be in the plane spanned by ${\bm c}(t)$, ${\bm c}_1(t)$, making them linearly independent: no solutions of (\ref{prima})-(\ref{terza}) can be found in this case.

\subsection*{Condition  {\bf 2)}} 
Consider next the case of condition {\bf 2)}:
since the  state $\rho(t)$ and the costate $\pi'(t)$ obey to the same evolution, once their Bloch vectors $\bm{a}(t)$ and $\bm{b}(t)$ become parallel, they will continue to be parallel for all the remaining time of the protocol. This means that we can equivalently check the condition at the final time $\tau$, rewriting it as 
\begin{equation}\label{cond12} 
\bm{a}(\tau) \parallel\bm{b}(\tau) = (0,0,-1)\;,
\end{equation} 
where we used the fact that 
 $\pi'(\tau)= -H_0$. This implies that condition {\bf 2)} can only be realized if 
 $\bm{a}(\tau)=\pm \lvert \bm{a}\rvert\ (0,0,1)$, 
What we have proved is that, to be in a singular interval, the state has to reach  either the minimum energy  achievable with a unitary evolution or the maximum one.
The first option is certainly unpleasant for an optimal control method, since it does not lead to an optimal protocol and for this reason we discard it. However, the second option would surely be the best protocol.

\subsection{Singular Interval analysis for the time-optimization problem}\label{singularapp} 

As seen in Sec.~\ref{sec:optimalchargingtime}, when optimizing the charging time for fixed final state $\rho_{\diamond}$ the function $G_1(t)$ has the same structure of the maximum energy optimization problem, see Eq.~(\ref{Eq:G0_def1}), the only difference being with the specific values of the vectors $\bm{a}(t)$ and 
$\bm{b}(t)$ which arise from dynamical equations which in principle are different from those of Eq.~(\ref{Eq:G0_condition}). Imposing the singular interval condition
$G_1(t)$ we hence get the same two possibilities detailed at the beginning of Sec.~\ref{appendix:qubit}.
Condition {\bf 1)} leads exactly to the identification of the same condition~(\ref{CONDsingular}), indeed also here we can rely on the fact that both $\bm{a}(t)$ and $\bm{b}(t)$ rotate around a common axis ${\bm n}^\star(t)$.

Condition {\bf 2)} requires however an independent analysis as now (\ref{cond12}) does not hold. 
Instead we can invoke the constraint~(\ref{newextra}) which expressed in terms of the controls of DCP problem becomes 
\begin{equation}
-i =\langle\pi(\tau)[H(\tau),\rho_{\diamond}]\rangle =
\frac{1}{4} \langle H(\tau)[\bm{a}_{\diamond} \cdot \bm{\sigma},\bm{b}(\tau)\cdot\bm{\sigma}]\rangle\;, 
\label{contofinmintime1}
\end{equation}
with $\bm{a}_{\diamond}$ being the Bloch vector of the target state $\rho_{\diamond}$. 
Observe next that the following identity applies 
\begin{eqnarray} \nonumber 
[\bm{a}_{\diamond} \cdot \bm{\sigma},\bm{b}(\tau)\cdot\bm{\sigma}] &=& 
[\bm{a}(\tau)  \cdot \bm{\sigma},\bm{b}(\tau)\cdot\bm{\sigma}] \nonumber \\
&=&-  4 [\rho(\tau), \pi(\tau)] \\ 
&=&-  4 U^{\star}_{\tau}(U^{\star}_t)^\dag [\rho(t), \pi(t)]  
 U^{\star}_t (U^{\star}_{\tau})^\dag\nonumber \\
 &=& U^{\star}_{\tau}(U^{\star}_t)^\dag [\bm{a}(t)  \cdot \bm{\sigma},\bm{b}(t)\cdot\bm{\sigma}] U^{\star}_t (U^{\star}_{\tau})^\dag \;, \nonumber  \label{diamond1}
\end{eqnarray} 
for all $t\in[0,\tau]$ and where we defined $U^{\star}_t := {\cal T}\exp[-i \int_0^t dt' H^{\star}(t')]$. The first of equalities \eqref{diamond1} is a consequence of the constraint $\rho(\tau)=\rho_{\diamond}$, the second  and the fourth derive
from the Bloch representation of the state and of the costate, the third from the
unitarity of the evolution. It is hence clear that if we do have a case where $\bm{a}(t)$ is parallel to 
$\bm{b}(t)$ for some time $t$, then $[\bm{a}(t) \cdot \bm{\sigma},\bm{b}(t)\cdot\bm{\sigma}]=0$ leading to 
a contradiction when replaced into (\ref{contofinmintime1}). 
This means that  for the time optimization problem, enforcing condition {\bf 2)} to identify the presence of singular
time intervals
 always leads to a contradiction:   Eq.~(\ref{CONDsingular}) is the only option that we have. 

\section{PMP analysis for the  qubit DCP with  two charging fields}
\label{newapp} 
Here we show that the solutions (\ref{solu1}) and (\ref{solu2}) fulfil the PMP condition~(\ref{condPMP}). 

Let us start by considering first the case (\ref{solu2}) where during the entire charging interval  $\tilde{\bm{\lambda}}^{\star}(t)$ maintains a constant value equal
to $r_{\max} \hat{\bm k}$. 
By direct integration of Eq.~(\ref{equest0})  we get
\begin{equation}\label{defbo} 
\tilde{\bm b}(t) = {\bm b}_{\|}(0)  + {\bm b}_{\bot}(0) \cos (2 r_{\max} t) + ( \hat{\bm k}\wedge {\bm b}_{\bot}(0) ) \sin ( 2r_{\max} t)\;,
\end{equation} 
with ${\bm b}_{\|}(0)$ and ${\bm b}_{\bot}(0)$  the components of ${\bm b}(0)$ which are parallel and orthogonal to  $\hat{\bm k}$, respectively. 
Notice however that since  $\tilde{\pi}'(\tau)= - H_0$, we must have  
$\tilde{{\bm b}}(\tau) = (0,0,-1) = - \hat{\bm x}_3$: replacing this into~(\ref{defbo}) and remembering that $\hat{\bm k}$ is orthogonal to  $\hat{\bm x}_3$ (see Eq.~(\ref{defK})), we can  conclude that ${\bm b}_{\|}(0)  =0$. Hence Eq.~(\ref{defbo}) simplifies to
\begin{eqnarray}\label{defbonew} 
&&\tilde{\bm b}(t) =  {\bm b}(0) \cos (2 r_{\max} t) + ( \hat{\bm k}\wedge {\bm b}(0) ) \sin ( 2r_{\max} t)~,\\
&&= \!|{\bm b}(0)|\! \!\left[ \cos( 2 r_{\max} t + \beta_0) \hat{\bm x}_3 \!-\! \sin( 2 r_{\max} t \!+ \!\beta_0) 
 ( \hat{\bm x}_3 \! \wedge \!\hat{\bm k}  )\right] \nonumber \!, 
\end{eqnarray} 
with 
\begin{equation} \beta_0 := \arccos\left(\frac{{\bm b}(0) \cdot \hat{\bm x}_3}{|{\bm b}(0)|}\right)=\arccos b_3(0)\in [0,\pi]\;. \end{equation} 
Comparing Eq.~(\ref{defbonew})
with (\ref{defao}) reveals that for the entire dynamical evolution 
$\tilde{\bm a}(t)$ and $\tilde{\bm b}(t)$ lay on the plane orthogonal to $\hat{\bm k}$, rotating
with the same constant angular velocity given by $r_{\max}$.  In particular, this implies that 
their vectorial product is constant in time during the entire evolution and  pointing into a direction which is anti-parallel to the rotation axis  $\hat{\bm k}$, i.e. 
\begin{eqnarray}\label{defbonew1} 
\tilde{\bm b}(t) \wedge \tilde{\bm a}(t) &=& 
 \tilde{\bm b}(\tau) \wedge \tilde{\bm a}(\tau) =  - \hat{\bm x}_3  \wedge \tilde{\bm a}(\tau)
 \\  \nonumber 
 &=&-  |{\bm a}(0)|  \sin( 2r_{\max} \tau + \alpha_0) 
 \;  \hat{\bm k} \nonumber \;, \end{eqnarray} 
 where we use~(\ref{defao}) and the fact that $\hat{\bm x}_3  \wedge(\hat{\bm x}_3  \wedge
  \hat{\bm k}) = - \hat{\bm k}$ [Notice that since $\tau\leq \tau_1$ we have that 
  $2r_{\max} \tau + \alpha_0 \leq \pi$ so that 
  $\sin( 2r_{\max} \tau + \alpha_0) \geq 0$]. 
From this Eq.~(\ref{condPMP}) now  follows by observing that 
\begin{eqnarray}\label{condPMP1}
  - \tilde{{\bm{\lambda}}}(t)  \cdot \tilde{\bm{b}}(t)\wedge\tilde{\bm{a}}(t) &\leq&
  |\tilde{{\bm{\lambda}}}(t)| |\tilde{\bm{b}}(t)\wedge\tilde{\bm{a}}(t)| 
  \nonumber \\
&\leq&   r_{\max} |{\bm a}(0)|  \sin( 2r_{\max} \tau + \alpha_0) \nonumber \\
&=& - \tilde{\bm{\lambda}}^{\star}(t)  \cdot \tilde{\bm{b}}(t)\wedge\tilde{\bm{a}}(t)\;. \end{eqnarray} 

In the case described by Eq.~(\ref{solu1}) we are supposed to keep 
 $\tilde{\bm{\lambda}}^{\star}(t)$ equal to
 $r_{\max} \hat{\bm k}$ for all $t\in[0,\tau_1]$ and then to switch-off the control.
This means that for all $t\in]\tau_1,\tau]$ both $\tilde{\bm a}(t)$ and $\tilde{\bm b}(t)$ 
are constant and equal to their final values, i.e. 
\begin{eqnarray}\tilde{\bm a}(t)&=&\tilde{\bm a}(\tau) =-|{\bm a}(0)| \hat{\bm x}_3\;, \nonumber \\
\tilde{\bm b}(t)&=&\tilde{\bm b}(\tau)=- \hat{\bm x}_3\;.
\end{eqnarray} 
In particular this implies that they are parallel and this condition is also maintained in the initial part of the dynamics as they rotate around the same axis. Therefore in this case 
\begin{eqnarray}  \tilde{\bm{b}}(t)\wedge\tilde{\bm{a}}(t)=0\quad  \Longrightarrow \quad 
{\tilde{\bm G}}(t)=0\;, 
\end{eqnarray}
making the entire trajectory a  singular interval (hence satisfying~(\ref{condPMP})). 

\section{Singular Intervals for Harmonic Oscillator DPC model}
\label{appendix:harmonic_oscillator}
Here we study the presence of 
singular intervals for Harmonic Oscillator DPC model, i.e.  time intervals during which
the function $G_1(t)$ of Eq.~(\ref{defg1}) gets equal to zero. 
The fundamental observation is that 
in order to fulfil such constraint   it is necessary  to have
not just $G_1(t)=0$, but also $\frac{d^nG_1(t)}{dt^n}=0$ $\forall n$. 
Recalling Eq.~(\ref{defg1}) this implies 
\begin{eqnarray}\nonumber 
p_2(t)&=&2v_2(t)\,\omega_0\;, \\ 
 \frac{d^np_2(t)}{dt^n}&=& 2 \frac{d^nv_2(t)}{dt^n}\,\omega_0\;.
 \label{condizderiv}
\end{eqnarray}
By imposing the PMP conditions for optimality  in \eqref{pseudohamgen}, we have that the costates
of the Harmonic Oscillator DMP model evolve in the following way:
\begin{equation}
\begin{cases}
\dot{p}_1(t)=0\;,\\
\dot{p}_2(t)= -\omega_0(2\lambda_1(t) + p_3(t)) +2 p_1(t) \lambda_1(t)\;,\\
\dot{p}_3(t)= \omega_0 p_2(t)\;,
\end{cases}
\label{costateevol}
\end{equation}
with boundary condition $p_j(\tau)=0$ for all $j$ [Notice that in particular this already
tells us that $p_1(t)=0$ for all $t$ so that we can eliminate it from the list].

From Eq.\ \eqref{condizderiv} with $n=1$ and from Eqs.\ \eqref{motostati} and \eqref{costateevol} we have:
\begin{eqnarray} \nonumber 
{\dot{{p}}_2(t)}&=&2 \omega_0 \dot{v}_2(t)=-2\omega^2_0 v_3(t) -2 
\omega_0 \lambda_1(t)\\
&=&-\omega_0 {p}_3(t) -2 \omega_0 \lambda_1(t)\end{eqnarray}
from which we get 
\begin{eqnarray} 
{p}_3(t)&=&2 \omega_0 v_3(t)\;. 
\end{eqnarray}
In conclusion the conditions to be in a singular interval are
\begin{equation}
{p}_2(t)=2 \omega_0 v_2(t)\;, \qquad 
{p}_3(t)=2 \omega_0 v_3(t)\;.
\label{singularosc}
\end{equation}
Now since up to a constant rescaling ${p}_2(t)$ and ${p}_3(t)$ have the exact same evolution of $v_2(t)$ and $v_3(t)$ respectively, it is evident that if \eqref{singularosc} holds at a time $t$, then it will continue to be true until $t=\tau$.
From the boundary conditions ${\bm{p}}(\tau)=0$, we obtain that $v_2(\tau)=v_3(\tau)=0$, i.e.\
\begin{equation}
\langle a \rho(\tau)\rangle=0\;. 
\end{equation} 
Notice however that if we assume that the input state of the system is the ground state of $H_0$, then the above condition
can only verified iff $\rho(\tau)$ corresponds to the ground state itself (a condition that is certainly unpleasant for an optimal control method that aims to increase the energy of the model). This fact follows from the observation that 
the Hamiltonian  \eqref{hamosc} can only induces displacements or phase shifts in the system, so that starting
from the vacuum it will always produce coherent states. And the only coherent state that has zero expectation value for
the annihilation operator is indeed the vacuum itself. 
This means that it can not  exists an optimal $\lambda_1^{\star}(t)$ that could enforce  condition for singularity.

\section{Harmonic Oscillator Frequency Optimization}
\label{app:oscopt}
Here we show that, among all the the Bang-Bang solutions that are optimal according to the results presented in section \ref{sec:harmoniccharge}, a square wave with a resonant frequency achieve the best performance in the long time limit.
From the dynamical equations for $v_2$ and $v_3$ in $\eqref{motostati}$ we obtain 
\begin{equation}
  \ddot{v}_3(t) + \omega_0^2 v_3(t) = - \omega_0 \lambda_1(t).  \label{eq:harmclass}
\end{equation}
The differential equation above can be solved using the Green's function approach.
The retarded Green's function satifsying $[\frac{d^2}{dt^2} + \omega^2] G(t-t')= \delta(t-t')$ can be computed with the Fourier Transform and reads
\begin{equation}
G(t-t') = \frac{1}{2 \pi}\int_{-\infty}^{\infty} \frac{e^{- i \omega (t-t')}}{\omega^2 + i \omega \epsilon -\omega_0^2}  d\omega 
\end{equation}
where $\epsilon > 0$ is a small parameter that we will send to $0$ at the end of the calculations.
The general solution of \eqref{eq:harmclass} is
 \begin{equation} 
 v_3(t) = v_3(0) \cos(\omega_0 t) + 
 \frac{\dot{v}_3(0)}{\sin(\omega_0 t)} 
 - \int_{0}^{t} dt'  \omega_0 G(t-t') \lambda_1(t')
 \end{equation}
that by initializing the battery in the ground state, i.e. by choosing $v_3(0)=\dot{v}_3(0) =0$,  reduces to
\begin{equation}
v_3(t) = \frac{1}{2 \pi}\int_{-\infty}^{\infty} dt'\int_{-\infty}^{\infty} \frac{\omega_0 e^{- i \omega (t-t')}}{\omega^2 + i \omega \epsilon -\omega_0^2} \lambda_1(t')  d\omega \;.
\end{equation}
Combining the results above with the first and the last of equations \eqref{motostati}, we obtain that the charging power writes 
\begin{equation}
v_1(t) \!=\! \frac{i}{\pi}\!\int_{-\infty}^{\infty}\! \!\!ds \int_{-\infty}^{\infty} \!\!dt'\!\!\int_{-\infty}^{\infty} \frac{\omega e^{- i \omega (s-t')}}{\omega^2\! + \!i \omega \epsilon -\omega_0^2} \lambda_1(t') \lambda_1(s)  d\omega\;,
\end{equation}
where the last integral is non zero only in the interval $[0,\tau]$, that is, when the external driving force is switched on. 
After performing the integrals on the time variables we are left with
 \begin{equation}
 v_1(t) =\frac{i \omega_0}{\pi}\int_{-\infty}^{\infty} \frac{\omega |\lambda_1(\omega)|^2}{\omega^2 + i \omega \epsilon -\omega_0^2}  d\omega\;.
 \end{equation}
With the residue theorem, after sending $\epsilon$ to $0$, the previous integral gives
\begin{equation}
    v_1(t) \approx c (|\lambda_1(\omega_0)|^2 + |\lambda_1(-\omega_0)|^2)\;,
\end{equation}
 where $c$ is a constant and the time dependence of $v_1$ is hidden in the parametric dependence of $\lambda_1(\omega)$ by time (we remember that the control has to nullify outside $[0,t]$).
 From the equation above, we obtain that to maximize the total work in the long time limit we have to choose a protocol that maximizes $|\lambda_1(\omega_0)|^2 + |\lambda_1(-\omega_0)|^2$.
 In the set of Bang-Bang protocols that we proved to be optimal, the best choice is a square wave with frequency $\omega_0$.

\bibliographystyle{myieeetr}

\end{document}